\documentclass[12pt,letterpaper]{article}
\pdfoutput=1

\usepackage{amsmath,amssymb,calc, amsthm,bbm}
\usepackage[utf8]{inputenc}
\usepackage{graphicx}
\usepackage{subfig}
\usepackage{hyperref}
\usepackage[nosort]{cite}
\usepackage{epsfig,psfrag}
\usepackage{slashed}
\usepackage{booktabs}

\makeatletter
\renewcommand\section{\@startsection {section}{1}{\z@}%
                                   {-3.5ex \@plus -1ex \@minus -.2ex}
                                   {2.3ex \@plus.2ex}%
                                   {\normalfont\large\bfseries}}
\renewcommand\subsection{\@startsection{subsection}{2}{\z@}%
                                     {-3.25ex\@plus -1ex \@minus -.2ex}%
                                     {1.5ex \@plus .2ex}%
                                     {\normalfont\bfseries}}
\makeatother

\theoremstyle{plain}

\theoremstyle{definition}

\let\non\nonumber

\let\a=\alpha
\let\b=\beta

\let\s=\sigma

\let\S=\Sigma

\newcommand{\del}{\partial}

\def\one{^{(1)}}

\newcommand{\bea}{\begin{eqnarray}}
\newcommand{\eea}{\end{eqnarray}}
\newcommand{\be}{\begin{equation}}
\newcommand{\ee}{\end{equation}}
\newcommand{\bma}{\begin{pmatrix}}
\newcommand{\ema}{\end{pmatrix}}

\newcommand{\Z}{{\mathbb Z}}
\newcommand{\R}{{\mathbb R}}

\newcommand{\PP}{{\mathbb P}}
\newcommand{\CC}{{\mathbb C}}

\newcommand{\com}[2]{{ \left[ #1, #2 \right] }}

\newcommand{\p}{\partial}

\def\com#1#2{{ \left[ #1, #2 \right] }}

\newcommand{\C}[1]{$(\ref{#1})$}

\def\IZ{\relax\ifmmode\mathchoice
{\hbox{\cmss Z\kern-.4em Z}}{\hbox{\cmss Z\kern-.4em Z}}
{\lower.9pt\hbox{\cmsss Z\kern-.4em Z}} {\lower1.2pt\hbox{\cmsss
Z\kern-.4em Z}}\else{\cmss Z\kern-.4em Z}\fi}
\def\IR{\relax{\rm I\kern-.18em R}}

\def\one{{\hbox{ 1\kern-.8mm l}}}

\newlength{\bredde}
\def\slash#1{\settowidth{\bredde}{$#1$}\ifmmode\,\raisebox{.15ex}{/}
\hspace*{-\bredde} #1\else$\,\raisebox{.15ex}{/}\hspace*{-\bredde}
#1$\fi}

\newsavebox{\zzzbar}
\sbox{\zzzbar}
  {\setlength{\unitlength}{0.9em}
  \begin{picture}(0.6,0.7)
  \thinlines
  \put(0,0){\line(1,0){0.6}}
  \put(0,0.75){\line(1,0){0.575}}
  \multiput(0,0)(0.0125,0.025){30}{\rule{0.3pt}{0.3pt}}
  \multiput(0.2,0)(0.0125,0.025){30}{\rule{0.3pt}{0.3pt}}
  \put(0,0.75){\line(0,-1){0.15}}
  \put(0.015,0.75){\line(0,-1){0.1}}
  \put(0.03,0.75){\line(0,-1){0.075}}
  \put(0.045,0.75){\line(0,-1){0.05}}
  \put(0.05,0.75){\line(0,-1){0.025}}
  \put(0.6,0){\line(0,1){0.15}}
  \put(0.585,0){\line(0,1){0.1}}
  \put(0.57,0){\line(0,1){0.075}}
  \put(0.555,0){\line(0,1){0.05}}
  \put(0.55,0){\line(0,1){0.025}}
  \end{picture}}

\newcommand{\ena}{\end{eqnarray}}
\newcommand{\beqa}{\begin{eqnarray}}
\newcommand{\eeqa}{\end{eqnarray}}

\newcommand{\half}{\frac{1}{2}}

\def\bes #1\ees{\begin{split}#1\end{split}}

\renewcommand{\b}{\beta}

\newfont{\goth}{ygoth.tfm scaled 1200}                   

\def\a{\alpha}
\def\b{\beta}

\def\th{\theta}

\def\s{\sigma}

\def\S{\Sigma}

 \numberwithin{equation}{section}

\def\1{{(1)}}
\def\2{{(2)}}
\def\3{{(3)}}

\def\1{{\bf 1}}
\def\a{{\alpha}}

\def\M{{\mathcal M}}

\def\CC{{\mathbb C}}

\def\1{{\bf 1}}
\def\3{{\bf 3}}
\def\7{{\bf 7}}
\def\2{{\bf 2}}
\def\8{{\bf 8}}

\newcommand{\h}[1]{{\widehat{#1}}}

\DeclareMathOperator{\rept}{Re}
\DeclareMathOperator{\impt}{Im}

\def\bes #1\ees{\begin{split}#1\end{split}}

\begin{document}
\begin{titlepage}

\begin{center}

October 2, 2018
\hfill         \phantom{xxx}  EFI-18-14

\vskip 2 cm {\Large \bf (2,2) Geometry from Gauge Theory} 
\vskip 1.25 cm {\bf Jo\~ao Caldeira$^{a}$, Travis Maxfield$^{b}$, and Savdeep Sethi$^{a}$}\non\\
\vskip 0.2 cm
 $^{a}$ {\it Enrico Fermi Institute \& Kadanoff Center for Theoretical Physics \\ University of Chicago, Chicago, IL 60637, USA}
\vskip 0.2 cm
$^{b}${\it {Center for Geometry and Theoretical Physics, Box 90318 \\ Duke University, Durham, NC 27708, USA}
}

\vskip 0.2 cm
{ Email:} \href{mailto:jcaldeira@uchicago.edu}{jcaldeira@uchicago.edu}, \href{mailto:travis.maxfield@gmail.com}{travis.maxfield@duke.edu}, \href{mailto:sethi@uchicago.edu}{sethi@uchicago.edu}

\end{center}
\vskip 1 cm

\begin{abstract}
\baselineskip=18pt




Using gauge theory, we describe how to construct 
generalized K\"ahler geometries with $(2,2)$ two-dimensional supersymmetry, which
are analogues of familiar examples like projective spaces and Calabi-Yau manifolds.  
For special cases, T-dual descriptions can be found which are squashed K\"ahler spaces. 
We explore the vacuum structure of these gauge theories by studying the Coulomb branch, which usually encodes the quantum cohomology ring. Some models without K\"ahler dual descriptions possess unusual Coulomb branches. Specifically, there appear to be an infinite number of supersymmetric vacua.

\end{abstract}

\end{titlepage}

\tableofcontents

\section{Introduction} \label{intro}

This work concerns sigma models in two dimensions with target space $\M$ and local coordinates $\phi$. Ignoring fermions, the bosonic sigma model action in a  target space patch takes the form
\be \label{sigmamodel}
S = \frac{1}{4\pi \alpha'} \int d^2x \sqrt{h} h^{\a\b}G_{ij} \p_\a \phi^i \p_\b \phi^j + i \int \phi^*(B),
\ee
where $ h$ is the two-dimensional worldsheet metric, while $G$ and $B$ denote the target space metric and $2$-form $B$-field.  Requiring extended worldsheet supersymmetry means introducing fermions and also restricting the target space $\M$. Of particular interest are models with chiral $(0,2)$ worldsheet supersymmetry, suitable for the heterotic string, and models with non-chiral $(2,2)$ supersymmetry suitable for both the heterotic and type II strings. 

Target spaces which are compatible with $(2,2)$ supersymmetry are called generalized K\"ahler spaces~\cite{Gates:1984nk, Hitchin:2004ut, Gualtieri:2014kja}. There are two basic issues one might try to address. The first is classifying the geometric structures required for $\M$ to admit $(2,2)$ supersymmetry, and the corresponding implications for superspace constructions. There has been a great deal of progress along these lines starting with~\cite{Gates:1984nk}. For general $(2,2)$ non-linear sigma models, the basic needed superspace ingredients are chiral, twisted chiral and semi-chiral superfields~\cite{Lindstrom:2005zr}. See~\cite{Bischoff:2018kzk}\ for a recent discussion of the defining data for classes of $(2,2)$ models. 

The second issue is the question of constructing classes of $(2,2)$ target spaces. This question has a somewhat different flavor because acceptable target spaces can include ingredients that require a physical explanation; for example, spaces with orbifold singularities, particularly those with discrete torsion, brane sources, or the use of stringy worldsheet symmetries like T-duality in patching conditions. 

The simplest examples of $(2,2)$ sigma models have K\"ahler target spaces $\M$. Imposing conformal invariance further restricts $\M$ to a Calabi-Yau space. Once one reaches Calabi-Yau $4$-folds, there are believed to be an enormous number of such spaces with lower bound estimates of $O(10^{755})$~\cite{Halverson:2017ffz}, and a recent Monte-Carlo based estimate of $O(10^{3000})$~\cite{Taylor:2017yqr}! On top of this geometric degeneracy is the usual enormous number of choices of flux, estimated in one case to be $O(10^{272,000})$~\cite{Taylor:2015xtz}. Somewhat surprising is the realization that a large fraction of these Calabi-Yau spaces admit elliptic fibrations and even $K3$-fibrations~\cite{Anderson:2016cdu}.     

Duality between the heterotic string and $K3$-fibered F-theory flux vacua, built from Calabi-Yau $4$-folds, suggests that there should exist an enormous number of worldsheet string geometries with non-vanishing $H=dB$~\cite{Dasgupta:1999ss}. These are not Calabi-Yau manifolds but rather a kind of torsional background compatible with $(0,2)$ worldsheet supersymmetry.  The expected number of such geometries should dwarf the number of currently known Calabi-Yau $3$-folds. Yet very few compact examples are known. Unlike the case of K\"ahler target spaces, there are few if any systematic constructions of flux geometries with $H \neq 0$. We are missing tools like algebraic geometry which might provide us with large classes of such spaces. 

This picture motivates us to move away from the familiar K\"ahler geometries visible under the lamp post, and search for the new ingredients and structures needed to describe more generic string geometries with non-vanishing $H$. Along the way, we will learn more about the physics of NS-branes and anti-branes. By an NS-brane we mean a localized magnetic source for $B$ such that the charge is non-vanishing, 
\be \label{nsbrane}
\int_{C_3} H \neq 0, 
\ee
where $C_3$ encloses the brane. The sign of the charge distinguishes a brane from an anti-brane. When the sigma model~\C{sigmamodel}\  is conformal and can serve as a classical string background, these NS-branes are the familiar NS5-branes. However, the definition~\C{nsbrane}\ applies to both gapped and conformal sigma models.

The approach we will take is to generalize the gauged linear sigma model (GLSM) construction described by Witten~\cite{Witten:1993yc}. Our generalization is motivated by the $(0,2)$ constructions described in~\cite{Adams:2006kb, Adams:2009av, Adams:2012sh, Quigley:2011pv, Melnikov:2012nm}, and specifically~\cite{Quigley:2012gq}. We will provide analogous constructions for models with $(2,2)$ supersymmetry. The enhanced $(2,2)$ supersymmetry makes a far larger set of tools available for analysis. 
While the most general model with $(2,2)$ supersymmetry involves semi-chiral superfields, in this work we will restrict our discussion to models constructed from chiral superfields $\Phi$ satisfying 
\be
\bar{D}_+ \Phi=\bar{D}_- \Phi= 0,
\ee
and twisted chiral superfields $Y$ satisfying
\be
\bar{D}_+ Y=D_-Y=0. 
\ee
Our conventions are described in Appendix~\ref{conventions}. We will also only consider abelian gauge theories. In the usual K\"ahler setting, this corresponds to considering toric spaces $\M$. Generalizing these constructions by considering non-abelian gauge theories, and by including semi-chiral representations is likely to be interesting.  

The main new ingredient over the original work of~\cite{Witten:1993yc}\ is the inclusion of periodic superfields,
\be\label{Yper}
Y \sim Y + 2\pi i. 
\ee
Such periodic fields appear in mirror descriptions of $(2,2)$ and $(0,2)$ GLSM theories~\cite{Hori:2000kt,Adams:2003zy, Melnikov:2012hk}, and in earlier GLSM constructions for torsional target spaces~\cite{Hori:2002cd,Tong:2002rq,Adams:2006kb}. The superfield $Y$ can be used to build field-dependent Fayet-Iliopoulos  (FI) couplings, 
\be\label{FI}
\int d^2x d\theta^+ d\bar{\theta}^- Y\Sigma, 
\ee
where $\S$ is the field strength for a vector superfield. This coupling leads to torsion in the target space. As we will see later, including more couplings respecting the periodicity~\C{Yper}, like twisted superpotentials involving $e^Y$, gives interesting physical models. 
%

\begin{figure}[ht]
\centering
\includegraphics[scale=0.45]{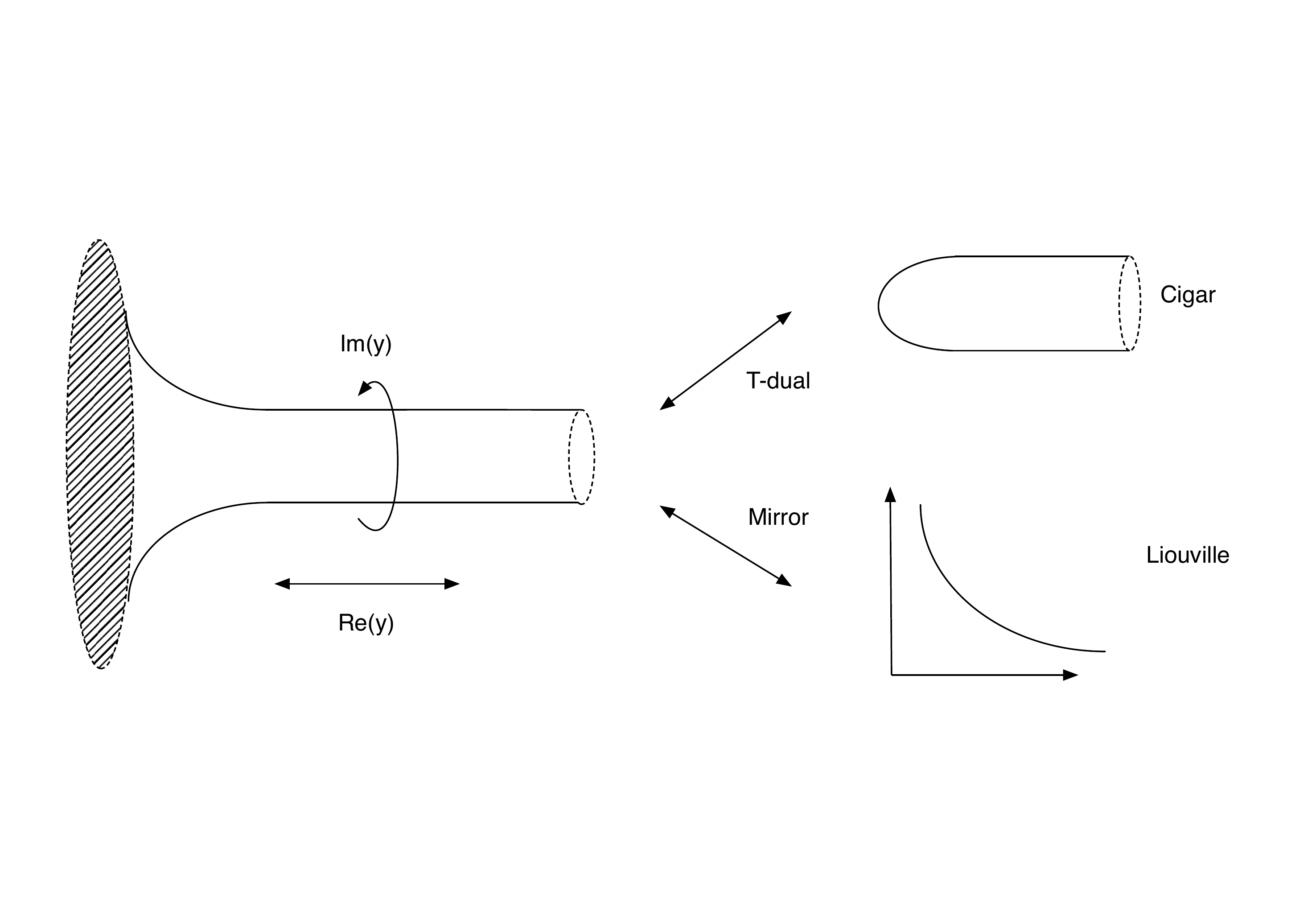}
\caption{\textit{The basic trumpet geometry for an NS-brane and its dual realizations. }}\label{fg}
\end{figure}

We can sketch the basic building blocks for our constructions. With one $U(1)$ gauge field and a field-dependent FI coupling~\C{FI}, the target space $\M$ is non-compact. 
This ingredient is depicted in Figure~\ref{fg}. The picture in $Y$ variables is the `trumpet' geometry, while a T-dual description gives the cigar geometry. Finally a mirror description gives the Liouville theory, which involves a potential energy coupling rather than pure geometry.  The blowing up of the ${\rm Im}(Y)$ circle is the hallmark of the brane in $Y$ variables. This  target space is a conformal field theory if one includes an appropriate varying dilaton field. The equivalence between the $(2,2)$ cigar and Liouville descriptions was argued by Giveon and Kutasov~\cite{Giveon:1999zm, Giveon:1999px}, building on earlier work and conjectures~\cite{fateev, Ooguri:1995wj}. A GLSM derivation of the equivalence was provided by Hori and Kapustin~\cite{Hori:2001ax}. The three pictures of the same physical system make clear the need to include T-duality -- the equivalence between large and small circles in string theory -- in the patching conditions for the target geometry $\M$. The trumpet is better described in terms of the cigar geometry near the locus where the circle blows up, while the cigar is better described by the Liouville theory when the asymptotic circle becomes small. 
In addition, the need for the Liouville description makes clear that we must allow potentials as well as metrics and $B$-fields when discussing more general notions of string geometry. This is quite reminiscent of the structure seen in hybrid Landau-Ginzburg phases; see, for example~\cite{Bertolini:2013xga, Chen:2018qww}.

\begin{figure}[ht!]
\centering
\includegraphics[scale=0.40]{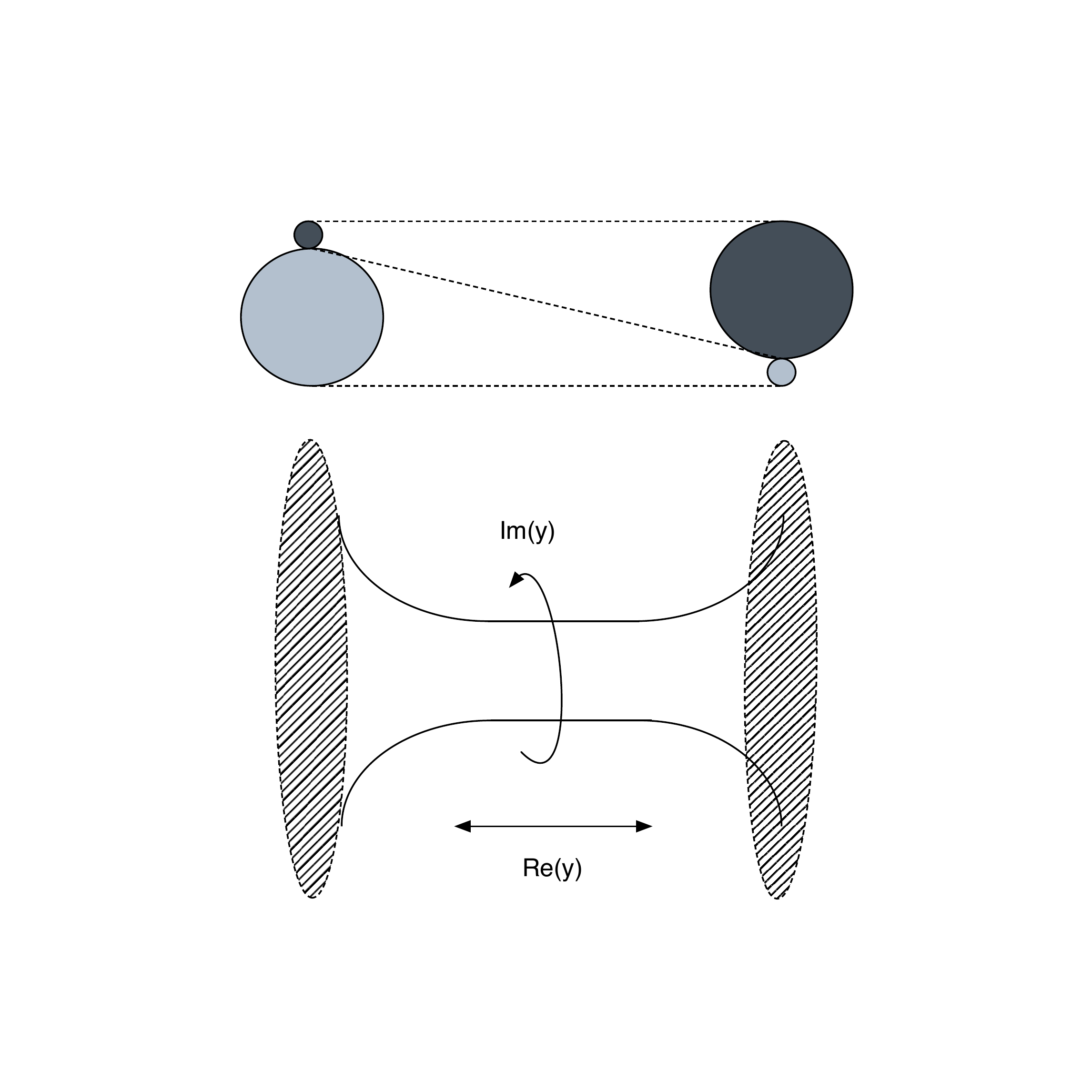}
\caption{\textit{A product of toric spaces fibered over the cylindrical fixture. }}\label{figure2}
\end{figure}

With multiple abelian gauge fields, the picture becomes richer. 
Instead of a semi-infinite trumpet, we can build finite-sized cylindrical fixtures. A product of toric varieties is typically fibered over each fixture with varying K\"ahler parameters. This is schematically depicted in Figure~\ref{figure2}. The interpretation of this geometry is that one end of the fixture supports a wrapped brane while the other end supports an anti-brane.  In a precise sense, these geometries are torsional dual descriptions of compact squashed toric varieties like projective spaces, introduced by Hori and Kapustin~\cite{Hori:2001ax}. This duality, which generalizes the standard relation between NS5-branes and ALF spaces, is described in section~\ref{dualdescriptions}. As we will see below, it is useful to have both descriptions of the same physical system in order to explore generalizations.


\begin{figure}[ht!]
\centering
\includegraphics[scale=0.40]{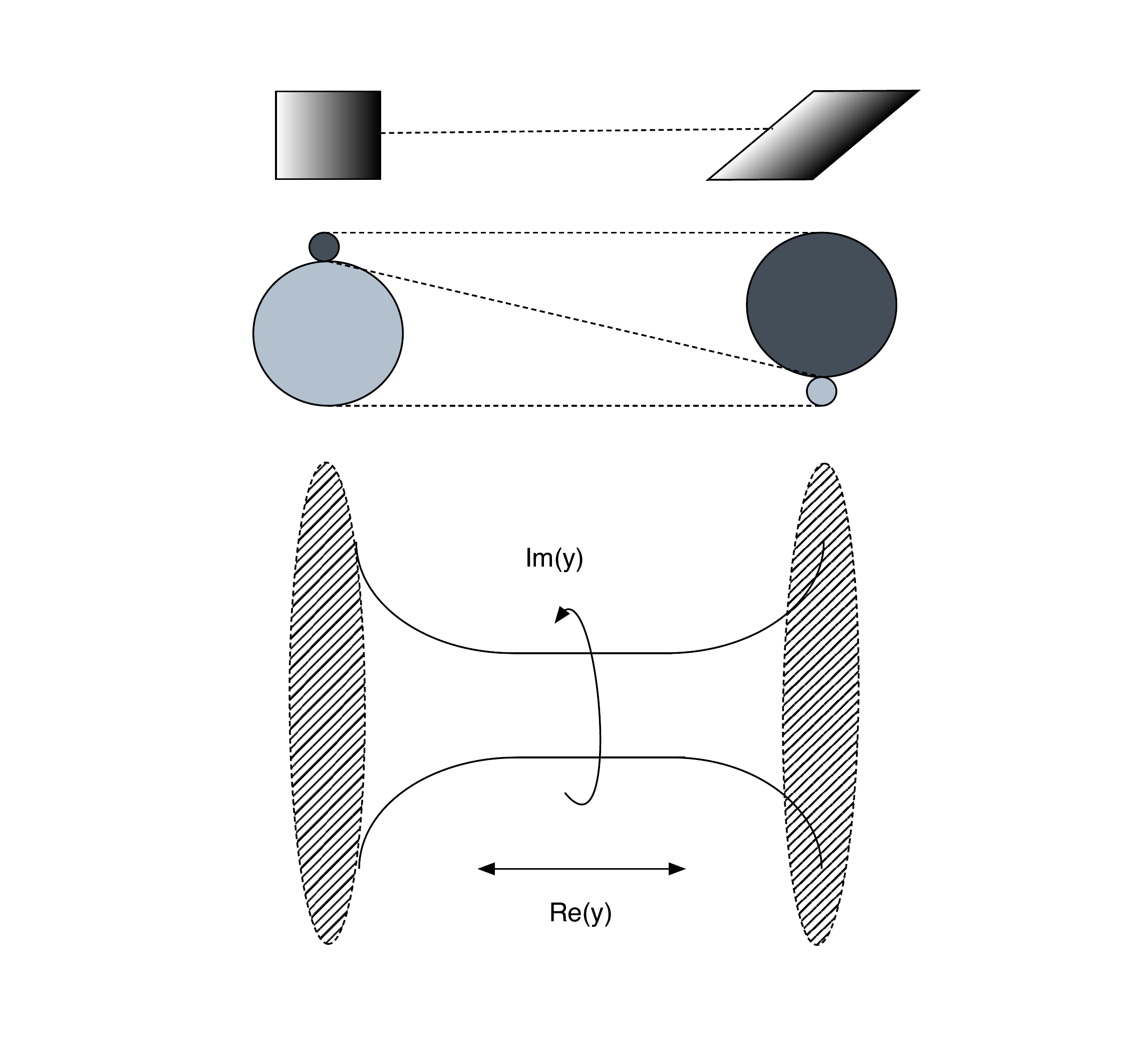}
\caption{\textit{The more general case with both complex and K\"ahler parameters fibered over the fixture. }}\label{figure3}
\end{figure}

Generic models involving this collection of ingredients cannot, however, be dualized to purely K\"ahler spaces; they are intrinsically torsional. We discuss two flavors of such models. The first flavor involves allowing the complex structure parameters of a fibered space to vary with $Y$. In a sense, this is the mirror version of the K\"ahler fibration of Figure~\ref{figure2}. This leads to generalized K\"ahler spaces of schematic form depicted in Figure~\ref{figure3}.  We describe examples of such models in section~\ref{examples}, and discuss the appearance of flat directions when one attempts to construct conformal models -- the analogues of Calabi-Yau spaces -- as complete intersections in the torsional analogues of toric varieties. We find it plausible that the flat directions we persistently find when a fibered toric space shrinks to zero size reflect the non-perturbative space-time physics supported on the NS-branes. This would match other examples where space-time non-perturbative physics, like the appearance of enhanced gauge symmetry at an ALE singularity, is reflected by a new branch in the associated gauged linear sigma model. 

The other generalization leads to wilder structures. Instead of simply allowing couplings like~\C{FI}\ which preserve the $U(1)$ isometry that shifts the imaginary part of $Y$, we consider more general field-dependent FI couplings, 
\be\label{generalFI}
\int d^2x d\theta^+ d\bar{\theta}^- f(Y, e^Y) \Sigma.  
\ee
The inclusion of interactions like this in otherwise topological interactions is familiar from $\mathcal{N}=1$ and $\mathcal{N}=2$ $D=4$ gauge theory, where superpotential or prepotential interactions can be generated by instantons or strong coupling effects. In $D=2$, we can include such couplings in the ultraviolet model and such models are described in section~\ref{sect:expysigma}. The classical vacuum equations have interesting new properties. We comment on some puzzling but interesting aspects of the quantum vacuum structure on the Coulomb branch of these models in section~\ref{sect:quantumcoh}. These Coulomb branch vacua are usually related to quantum cohomology rings of the target space $\M$. We find that the inclusion of these more general $U(1)$ breaking couplings leads to an infinite number of discrete Coulomb branch vacua, which is in sharp contrast to the finite number of vacua found in GLSMs describing K\"ahler spaces. We also find an analogue of the quantum cohomology ring for a class of generalized K\"ahler examples.  

Finally, it is worth noting that the way in which the spaces are constructed has a flavor similar to a recent construction of $G_2$ spaces at the level of geometry and conformal field theory~\cite{Corti:2012kd, Braun:2017uku, Fiset:2018huv}. It would be very interesting if this gauged linear construction can be generalized to produce target geometries with $G_2$ holonomy.  

\section{Dual Descriptions}\label{dualdescriptions}

We will consider models built from chiral superfields $\Phi$ with charge $Q$ under an abelian gauge group. All basic conventions are found in Appendix~\ref{basicconventions}. These superfields contain one complex scalar $\phi$. Under a gauge transformation with chiral superfield parameter $\Lambda$, the vector superfield $V$ for the gauge symmetry and $\Phi$ transform as follows:
\begin{align}
V &\rightarrow V+\frac{i}{2}(\bar{\Lambda}-\Lambda), \qquad \Phi\rightarrow e^{iQ\Lambda}\Phi. 
\end{align}
The field strength superfield defined in~\C{fieldstrength}\ is denoted $\S$. It contains one complex scalar $\s$. Following the notation of~\cite{Hori:2001ax}, we also consider chiral superfields $P$ which are shift charged under the gauge group:
\be P\rightarrow P + i\ell\Lambda, \qquad P\sim P+ 2\pi i. \label{eqn:pperiodicity}
\ee
with $\ell \in \mathbb{Z}$. The chiral superfield $P$ also contains one complex scalar $p$. We will write the kinetic terms of $P$ as
\begin{equation}
    S^P_{\text{kin}} = \frac{b}{32\pi} \int d^2x d^4\theta \left(P + \bar{P} + 2\ell V\right)^2,
\end{equation}
allowing an overall factor $b$. We could allow the periodicity to vary as well, introducing an additional real parameter $\delta$ and defining $P$ such that $P \sim P + 2\pi i \delta$. However, we can always scale $P$ to have $\delta=1$, absorbing the change into $b$ and $\ell$. This is what we will do in this work.
The last ingredient is twisted chiral neutral superfields $Y$ with periodicity given in~\C{Yper} containing a complex scalar $y$.

Now this is a highly asymmetric treatment of chiral versus twisted chiral superfields. Mirror symmetry exchanges the two kinds of constrained superfield, and it was recognized in very early attempts to prove mirror symmetry that a more symmetric treatment might prove helpful. Following the terminology of Morrison and Plesser~\cite{Morrison:1995yh}, we introduce superfields for a twisted GLSM. These fields are denoted by a `hat' so $\h{\Phi}$ is a {\it twisted chiral} superfield charged under a vector superfield $\h{V}$ with {\it chiral} field strength $\h{\S}$. The {\it chiral} superfields $\h{Y}$ are neutral and periodic, just like their twisted chiral cousins in \eqref{Yper},
\be\label{hatYper}
\h{Y} \sim \h{Y} + 2\pi i. 
\ee
These are similar to the fields $P$ in \eqref{eqn:pperiodicity}, except we will always take fields $Y$ to not transform under the gauge group while $P$ is shift-charged. As we will see in section~\ref{generalcase}, all these field types are very natural for describing general models.

\subsection{Holomorphic data and kinetic terms}

We can group the collection of fields into uncharged periodic fields $(Y, \h{Y})$, charged fields $(\Phi, \h{\Phi}, P, \h{P})$ and gauge fields $(V, \h{V})$ with field strengths $(\S, \h{\S})$. The data over which we have the most control under RG flow are holomorphic couplings. These include the superpotential, 
\be
S_W = \frac{1}{8\pi}\int d^2x d\theta^+ d\theta^- \left\{ -i\h{t}\h{\S} - \h{k}\h{\S}\h{Y} +  W(\Phi, P, e^\h{Y}, \h{\S}) + \text{c.c.} \right\},
\ee
where $\h{t}$ determines the FI parameter via~\C{FIrtheta}, $\h{k}$ is an integer in order to be compatible with the $\h{Y}$ periodicity, and $W$ is gauge-invariant and single-valued. There is an analogous twisted chiral superpotential with form, 
\be \label{eqn:twistedsuperpotential}
S_{\widetilde W} = \frac{1}{8\pi}\int d^2x d\theta^+ d\bar{\theta}^- \left\{-it \S - k\S Y +  \h{W}(\h{\Phi}, \h{P}, e^Y, \S) + \text{c.c.} \right\},
\ee
where again $k \in \Z$ and $\h{W}$ is single-valued.\footnote{Actually the superpotential and twisted superpotential need not be classically gauge-invariant if one includes $\int d^4\theta V \h{V}$ couplings, which appear in~\cite{Morrison:1995yh}.} It is clear from the dependence of the superpotential on $t$ and $Y$ that we could choose to absorb the FI parameters $t$ into a constant shift of the fields $Y$, without affecting the $Y$ kinetic terms. However, the FI parameters $t$ will typically be additively renormalized. While we could choose to absorb this renormalization into a shift of $Y$, for clarity we choose to always keep $t$ explicit in this work. The function $\h{W}$ may include terms such as the typical polynomials in the charged fields $\h{\Phi}$, polynomials including $\h{\Phi}$ as well as $e^{\h{P}}$ and $e^Y$, and other terms including only $Y$ and $\Sigma$ like $e^Y\Sigma$. We will briefly explore these possibilities in what follows.

These are the obvious holomorphic couplings, but there is actually more potential holomorphic data hidden in the kinetic terms because of the periodic fields $(Y, \h{Y})$. Let us focus on the $Y$ fields. A general kinetic term for $Y$ takes the form,
\be
S^Y_{\rm {kin}} = -\frac{1}{16\pi}\int d^2x d^4\theta \, \left(\bar{Y} f_1 + f_2 + {\rm c.c.} \right), 
\ee
where the $f_i$ are gauge-invariant functions of the superfields, and $f_2$ is single-valued. To be sensible under the periodic identification~\C{Yper}, $f_1$ must be annihilated by $\int d^4\th$. This constrains $f_1$ to be a holomorphic function of the twisted chiral superfields; $f_1$ can also depend either holomorphically or anti-holomorphically on the chiral superfields, so that $\bar{D}_+f_1=0$ or $D_-f_1=0$. This holomorphic data is likely to prove interesting for non-linear sigma models. 

In a linear theory, we are typically interested in kinetic terms that are quadratic in the fields. At the quadratic level, there are no direct couplings of chiral and twisted chiral fields so the $Y$ and $\h{Y}$ fields do not kinetically mix. If there are $n$ $Y$-fields then the choice of $Y$ kinetic term corresponds to a choice of metric and $B$-field for $T^n$ encoded in $k_{\mu\bar\nu}$
\be S^Y_{\rm {kin}} = -\frac{1}{16\pi}\int d^2x d^4\theta \, k_{\mu\bar\nu} Y_\mu \bar {Y}_{\bar \nu}. 
\ee
The classical moduli space is the familiar $n^2$-dimensional Narain moduli space, 
\be
\frac{O(n,n, \R)}{O(n)\times O(n)}. 
\ee
We will not worry about discrete identifications on the moduli space since those identifications are not generally preserved by interactions. A similar discussion applies to the $\h{Y}$ fields. 

The charged and uncharged fields also do not mix kinetically at the level of quadratic interactions. For a $\Phi$ field with charge $Q$, we simply assume canonical kinetic terms, 
\be 
S^{\Phi}_{\rm kin}  = \frac{1}{16\pi}\int d^2x d^4\theta \, \bar{\Phi} e^{2QV}  \Phi, 
\ee
whose component form appears in~\C{chiralkinetic}, and similarly for a charged $\h{\Phi}$ field.  

\subsection{Worldsheet duality} \label{sect:Tduality}

Here we present the duality dictionary we will subsequently use. More details are presented in Appendix~\ref{sect:Tdualityexpanded}. A $(2,2)$ chiral superfield $P$ with periodicity given in~\eqref{eqn:pperiodicity} can be axially charged, making its imaginary part a two-dimensional Stueckelberg field. The action is simply,
\begin{equation}
S=\frac{b}{32\pi}\int d^2xd^4\theta \left(P+\bar{P}+2V\right)^2,
\end{equation}
where $b$ is a constant that will later be interpreted as a squashing parameter for the models of section \ref{sect:kahlerpicture}. This particular theory has a dual description in terms of a $(2,2)$ twisted chiral superfield $Y$ with the same periodicity, and with action:
\begin{equation}
S_d=-\frac{1}{16\pi b}\int d^2x d^4\theta  \, \bar{Y}Y-\frac{1}{8\pi}\int d^2x d\theta^+ d\bar{\theta}^- \, Y\Sigma + \text{c.c.}.
\end{equation}
This duality requires the coupling between $Y$ and $\Sigma$ found in~\C{FI}, which will feature heavily in this work.

Alternatively, a $(2,2)$ chiral $\Phi$ parametrizing $\CC$ can also be dualized. The action for a charged chiral takes the form,
\begin{equation}
S=\frac{1}{16\pi}\int d^2x d^4\theta \,|\Phi|^2 e^{2V}, 
\end{equation}
and the dual $Y$ is also a twisted chiral superfield with periodicity~\eqref{Yper}, and with action:
\begin{align}
S_d = &-\frac{1}{16\pi}\int d^2xd^4\theta\, (Y+\bar{Y})\log(Y+\bar{Y})   \nonumber\\ &  - \left[ \frac{1}{8\pi}\int d^2x d\theta^+d\bar{\theta}^- \,\left(Y\Sigma + \mu e^{-Y}\right)+\text{c.c.} \right]. \label{eqn:dualchargedphi}
\end{align}

\subsection{Anomaly and conditions for conformality}

Given a $(2,2)$ gauged linear sigma model defined in the ultraviolet, it is usually a non-trivial issue to decide whether or not it flows to a non-trivial conformal field theory. One way to strengthen the case for a non-trivial infrared fixed point is to construct a candidate $\mathcal{N}=2$ superconformal algebra in the ultraviolet theory. This boils down to finding a non-anomalous right-moving R-symmetry current.
In usual $(2,2)$ gauged linear sigma models built using only chiral superfields, this is possible if the sum of the charges vanishes,
\begin{align}
\sum_i Q_{ia} = 0,
\end{align}
for all gauge symmetries. This means that the FI parameters of the theory will be invariant under renormalization group flow, and that the curvature two-form for the non-linear model found by symplectic quotient will be trivial in cohomology, so there is a Ricci-flat metric in the same K\"ahler class. However, this condition is modified once we add $P$ and $Y$ fields. In this section we will find the more general condition.

Let us start by defining our R-symmetries in the gauge theory. Under $U(1)_R$, $\theta^+$ has charge 1, while under $U(1)_L$, $\theta^-$ has charge 1. In a V-A basis, $q_{\theta^+}=(1,1)$, and $q_{\theta^-}=(1,-1)$. Note $d\theta$ transforms as $\theta^{-1}$, so a chiral superpotential should have $R$ transformation with charges $(2,0)$ in that basis, while a twisted superpotential would have charges $(0,2)$. R-invariance of the gauge field fixes $q_\Sigma=(0,2)$; in turn, this implies that the FI twisted superpotential $t \S$ has the correct R-charge. 
A chiral superfield with charge $Q$ causes an anomaly under $U(1)_A$, since $q_{\psi_+}-q_{\psi_-}=-2$. This corresponds to a variation of the effective action
\begin{align} \label{eqn:anomalychargedphi}
\delta S = \frac{\beta Q(-2)}{4\pi}\int d^2x \epsilon^{\mu\nu}F_{\mu\nu},
\end{align}
where $\beta$ is the transformation parameter. The same expression is valid for a twisted chiral with charge $\h{Q}$ under $U(1)_V$. We can see from~\eqref{eqn:dualchargedphi} and~\eqref{eqn:ysigmacomponents} that reproducing this anomaly fixes the transformation under $U(1)_A$ for a field $Y$ dual to a charged $\Phi$ to be a shift charge of $-2$, so $e^{-Y}$ in~\eqref{eqn:dualchargedphi} transforms like a twisted superpotential.

On the other hand, when $Y$ is not dual to a charged $\Phi$, its $U(1)_A$ transformation is not fixed. Defining it generally to have a shift charge $2\gamma$ under $U(1)_A$, or $\pm \gamma$ under $U(1)_{R,L}$, will cause the action \eqref{eqn:twistedsuperpotential} with $\tilde{W}=0$ to vary as
\begin{align} \label{eqn:anomalyysigma}
\delta S = \frac{\beta (2 k \gamma)}{4\pi}\int d^2x \epsilon^{\mu\nu}F_{\mu\nu}
\end{align}
from the coupling~\eqref{eqn:ysigmacomponents}. The updated condition for a non-anomalous $U(1)_R$ symmetry with $Y^\mu$ fields coupled to gauge fields $\Sigma^a$ with coefficients $k_{\mu a}$ is then obtained by requiring the sum of variations \eqref{eqn:anomalychargedphi} and \eqref{eqn:anomalyysigma} to vanish,
\begin{align}\label{eq:conformalityconditionnop}
\sum_i Q_{ia} - \sum_\mu \gamma_{\mu} k_{\mu a} = 0
\end{align}
for some charges $\gamma_\mu$.

What about our fields $P$ with Stueckelberg couplings? At first sight, unlike $Y$ they do not have a coupling that would let us compensate an anomalous variation with a field transformation. However, in many cases we can take a field $Y$ and dualize it into $P$, so the same condition must apply in both pictures. To understand how this works, note that postulating a transformation $y \rightarrow y + i \gamma\beta$ means we are modifying the axial R-symmetry current by 
\begin{align}
j_+ &= \dots - \frac{i \gamma}{2}\del_+ (y-\bar{y}), &
j_- &= \dots - \frac{i \gamma}{2}\del_- (y-\bar{y}).
\end{align} 
If we dualize $y$ into $p$, from~\eqref{eqn:ypdualitymap}, these current modifications become
\begin{align}
j_+ &= \dots - \frac{i \gamma}{2}D_+ (p-\bar{p}), &
j_- &= \dots + \frac{i \gamma}{2}D_- (p-\bar{p}).
\end{align}
These are the current improvement terms constructed in \cite{Hori:2001ax}. These terms do not give $p$ a variation under $U(1)_A$, but they modify the current conservation equation. In conclusion, for each field $p_\alpha$ with shift charges $\ell_{\alpha a}$ under the gauge symmetries, we will also have a parameter that modifies the condition for a non-anomalous R-symmetry into
\begin{equation}\label{eq:conformalitycondition}
\sum_i Q_{ia} -\sum_\alpha \gamma_\alpha\ell_{\alpha a}- \sum_\mu \gamma_\mu k_{\mu a} = 0.
\end{equation}
This is the condition we must work with in models with $Y^\mu \Sigma^a$ couplings. On the other hand, note that a superpotential term of the form $e^{Y^\mu}\Sigma$ precludes any transformation of that field $Y^\mu$ under $U(1)_R$, setting $\gamma_\mu=0$. This is because $\Sigma$ alone must have the correct transformation for a twisted superpotential as long as the FI coupling is non-trivial.

When we do expect a theory to flow to a non-trivial infrared fixed point, we can construct the protected, right-moving chiral algebra in the ultraviolet whose central charge, in particular, agrees with that of the infrared theory in the absence of accidents~\cite{Witten:1993jg}. We will construct the right-moving superconformal multiplet for a general class of models here.

A very general model constructed from the ingredients we have considered is the following:
\begin{align}\label{eq:generalPYmodel}
S &= \frac{1}{16\pi} \int d^2x d^4 \theta \left( |\Phi_i|^2e^{2Q_{ia}V_a} + \frac{b_\alpha}{2} \left( P_\alpha + \bar P_\alpha + 2 \ell_{\alpha a} V_a \right)^2 - \frac{1}{b_\mu}|Y_\mu|^2 - \frac{2\pi}{e_a^2}|\Sigma_a|^2 \right) \nonumber \\
&-\frac{i}{8\pi} \int d^2x d\theta^+d\bar{\theta}^- \left( t_a -i k_{\mu a} Y_\mu \right)\Sigma_a + \text{c.c.} \\
& + \frac{1}{8 \pi} \int d^2x d\theta^+d\theta^-~W\left(\Phi_i, e^{P_\alpha} \right) + \text{c.c.}. \nonumber
\end{align}
For the moment, let us ignore the superpotential term. This model has a protected, right-moving superconformal multiplet of the form:
\begin{align}
\mathcal{J}^{0}_{--} &= -\frac{1}{8\pi} D_- \left( e^{2Q_i \cdot V}\Phi_i \right) e^{-2Q_i \cdot V} \bar D_- \left( e^{2Q_i \cdot V} \bar \Phi_i \right) \nonumber \\
& \phantom{=} -\frac{b_\alpha}{8\pi} D_-\left(P_\alpha + \bar P_\alpha + 2\ell_\alpha \cdot V \right) \bar D_-\left(P_\alpha + \bar P_\alpha + 2\ell_\alpha \cdot V \right)\\
&\phantom{= } -\frac{1}{8\pi b_\mu}D_-\bar Y_\mu \bar D_- Y_\mu -\frac{1}{4 e_a^2} \Sigma_a \bar D_- D_- \bar \Sigma_a. \nonumber
\end{align}
This is the superconformal multiplet assigning $R$-charge $0$ to the $\Phi_i$ fields. When we include the superpotential, we will modify this multiplet by the addition of a flavor symmetry under which the $\Phi_i$ rotate.

Classically, $\mathcal{J}_{--}^0$ satisfies
\begin{equation}
\bar D_+ \mathcal{J}_{--}^0 = 0;
\end{equation}
however, there is a $1$-loop anomaly that modifies this to
\begin{equation}
\bar D_+ \mathcal{J}_{--}^0 = \frac{\gamma_a}{4\pi} \bar D_- \Sigma_a.
\end{equation}
Following~\cite{Hori:2001ax}, we determine the anomaly coefficient $\gamma_a$ by point-splitting, the details of which are found in Appendix~\ref{app:centralchargedetails}. The result is
\begin{equation}
\gamma_a = \sum_i Q_{ia}.
\end{equation}
As in~\cite{Hori:2001ax}, a modified current can be defined that is $1$-loop superconformal. Unlike in that case, we will also take advantage of the freedom to include the $Y_\mu$ fields in that modification. In particular, we use
\begin{equation}
\mathcal{J}_{--} = \mathcal{J}_{--}^0 + \frac{\gamma_\alpha}{8\pi} \com{\bar D_-}{D_-} \left( P_\alpha + \bar P_\alpha + 2 \ell_{\alpha a}V_a \right) +\frac{\gamma_\mu}{8\pi \sqrt{b_\mu}} \com{\bar D_-}{D_-} \left( Y_\mu + \bar Y_\mu \right),
\end{equation}
where $\gamma_\alpha$ and $\gamma_\mu$ are chosen such that
\begin{equation}
\gamma_a = \gamma_\alpha \ell_{\alpha a}+ \gamma_\mu k_{\mu a} .
\end{equation}
This is the same as \eqref{eq:conformalitycondition}.

Including the superpotential, $\mathcal{J}_{--}$ is no longer $\bar D_+$ closed. Instead,
\begin{equation}
\bar D_+ \mathcal{J}_{--} = \frac{1}{4\pi}\left(  \left(\mathcal{D}_- \Phi_i\right) \partial_i  + \left( \mathcal{D}_- P_\alpha + \gamma_\alpha D_- \right) \partial_\alpha \right) W\left(\Phi_i, e^{P_\alpha} \right).
\end{equation}
We can partially remedy this by assuming the quasi-homogeneity of $W$ and including in $\mathcal{J}_{--}$ the flavor symmetry under which each chiral field $\Phi_i$ rotates with the corresponding degree:
\begin{equation}
\mathcal{J}_{--} \to \mathcal{J}_{--} + \mathcal{F}_{--}, \qquad \mathcal{F}_{--} = \frac{\alpha_i}{8\pi} D_- \bar D_- \left( |\Phi_i|^2 e^{2 Q_i \cdot V} \right), \qquad \left( \alpha_i \Phi_i \partial_i + \beta_\alpha \partial_\alpha \right) W  = W.
\end{equation}
Then,
\begin{equation}
\bar D_+ \left( \mathcal{J}_{--} + \mathcal{F}_{--} \right) = \frac{1}{4\pi}\left( \beta_\alpha + \gamma_\alpha \right) D_- \partial_\alpha W.
\end{equation}

Assuming the above is zero, i.e. $\beta_\alpha + \gamma_\alpha = 0$ and we have a non-trivial infrared CFT, we can find its central charge from the supercurrent we have just written. Details of this calculation are found in Appendix~\ref{app:centralchargedetails}. The result is
\begin{equation}\label{centralcharge}
c = 3 \left( \sum_i \left( 1 - 2\alpha_i \right) - N_{U(1)} + N_P + N_Y + 2 \sum_\alpha \frac{\gamma_\alpha^2}{b_\alpha} + 2 \sum_\mu \frac{\gamma_\mu^2}{b_\mu} \right).
\end{equation}
The first two terms are familiar from standard GLSM model building. The next two terms, $N_P$ and $N_Y$ are the number of $P$ and $Y$ fields, respectively, while the last two terms are modifications due to the non-standard shift transformations required of the $P$ and $Y$ fields in order to cancel all anomalies. In the next section, we will comment on specific examples of models and their supercurrents and central charges, when possible.

\section{A Collection of Models} \label{examples}

We will now describe a series of examples of varying complexity that serve to illustrate some of the possible target geometries described by this construction. Here we are only concerned with the classical geometry that emerges from minimizing the potential energy for a given GLSM. This corresponds to solving the $D$-term and $F$-term conditions, and quotienting by the gauge group action. For the usual K\"ahler setting, solving the $D$-term conditions and quotienting by the gauge group action defines a toric variety via symplectic quotient. Further imposing $F$-term conditions gives an algebraic variety.    

\subsection{Bounding \texorpdfstring{$D$}{D}-terms} \label{sect:boundingdterms}

\subsubsection{One \texorpdfstring{$U(1)$}{U(1)} action}  \label{sect:ysigmaoneu1}

Turning on a single field-dependent FI-term, we work with the action
\begin{align} \label{eqn:kysigmacoupling}
S &= \frac{1}{16\pi} \int d^2x d^4 \theta \left( \sum_i |\Phi_i|^2e^{2Q_{i}V} -|Y|^2 - \frac{2\pi}{e^2}|\Sigma|^2 \right) \nonumber \\
& \hspace{10pt} -\frac{i}{8\pi} \int d^2x d\theta^+d\bar{\theta}^- \left( t -i k Y \right)\Sigma + \text{c.c.}
\end{align}
leading to a $D$-term potential imposing
\be\label{singledterm} 
\sum_i Q_i |\phi_i|^2 = r - {2 k} {\rm Re}(y). 
\ee
For simplicity, let us assume all $Q_i$ are positive, $r\geq 0$ and $k \geq 0$. After quotienting by $U(1)$, the $\phi^i$ configuration space is a weighted projective space with size determined by the right-hand side of~\C{singledterm}. This space is non-compact since ${\rm Re}(y)$ is only bounded from above,   
\be \label{range}
{\rm Re}(y) \leq \frac{r}{2 k}. 
\ee
When the inequality in~\C{range}\ is saturated, the weighted projective space collapses to zero size. Although no fields charged under the gauge symmetry have expectation values at this boundary, $\Sigma$ is still massive because of the $Y\Sigma$ coupling which contributes a $|\s|^2$ mass term to the physical potential. There is therefore no classical Coulomb branch emitting from $|\phi_i|^2=0$.  Just like its dual Stueckelberg field $P$, the field $Y$ gives a mass to the gauge field everywhere, so in the limit $e\rightarrow \infty $ the gauge field is not dynamical. 

The simplest example has $n$ chiral fields $\Phi^i$ of charge $1$ and one twisted chiral $Y$ with the coupling \eqref{eqn:kysigmacoupling} and a standard kinetic term. After carrying out the symplectic quotient, we obtain a metric describing $\mathbb{C}P^{n-1}$ parametrized by $n-1$ complex coordinates $z^i$ fibered over a cylinder parametrized by $y$,
\begin{align} \label{eqn:metriconeu1ycomplex}
ds^2&=R(y)\left(\frac{dz\cdot d\bar{z}}{1+|z|^2}-\frac{|\bar{z}\cdot dz|^2}{(1+|z|^2)^2}\right)+\left(1+\frac{k^2}{R(y)}\right)dyd\bar{y}, \\
R(y)&=r-2k\rept (y)\end{align}
and also a $B$-field
\begin{align} \label{bfieldoneu1y}
B= \frac{k \bar{z}\cdot dz \wedge d\bar{y}+kz\cdot d\bar{z}\wedge dy}{1+|z|^2}.
\end{align}
We can write $y=a+i\theta$, absorb $r$ into $a$, and define $\rho^2\equiv R=2ka$ to rewrite the metric in a different form
\begin{align}\label{metriconeu1y}
ds^2=\rho^2\left(\frac{dz\cdot d\bar{z}}{1+|z|^2}-\frac{|\bar{z}\cdot dz|^2}{(1+|z|^2)^2}\right)+\left(1+\frac{\rho^2}{k^2}\right)d\rho^2+\left(1+\frac{k^2}{\rho^2}\right)d\theta^2.
\end{align}
This form makes it clear that $\rho$ has range $(0, \infty)$, with the projective space pinching to zero size at $\rho = 0$, while the circle parametrized by $\theta$ becomes infinitely large at that end. This is the trumpet geometry of Figure~\ref{fg}. This geometry is singular with diverging curvature as $\rho$ tends to zero. For example, when $n = 2$, the Ricci scalar is
\begin{align}
R = \frac{4}{\rho^2} + \ldots \, \ \mathrm{as} \, \ \rho \to 0.
\end{align}
While the space is geometrically singular, the theory has no physical singularity; the resolution of the singularity requires T-duality and we will be discussed in section~\ref{sect:kahlerpicture}.  

We can also calculate
\begin{align}
H = dB =2 k d\theta \wedge J_{FS}
\end{align}
where $J_{FS}$ is the fundamental two-form of Fubini-Study for $\mathbb{C}P^{n-1}$, which integrates to a non-trivial torsion
\begin{align}
\int_{\mathcal{C}\times S^1} H = 4\pi k
\end{align}
where $\mathcal{C}$ is the two-cycle dual to $J_{FS}$, where the dual is taken at fixed $\rho$ in the $\mathbb{C}P^{n-1}$ fiber, and the $S^1$ is parametrized by $\theta$.

In this model we can pick an $R$-symmetry transformation for $Y$ such that $n-k\gamma = 0$, so that there is a non-anomalous $U(1)_A$, and we expect a non-trivial infrared CFT. Using~\C{centralcharge}, the central charge of this CFT is calculated to be
\begin{equation}
c = 3n \left( 1 + \frac{2 n}{k^2} \right).
\end{equation}

\subsubsection{Two \texorpdfstring{$U(1)$}{U(1)} actions}\label{sect:ysigmatwou1}

With at least one more $U(1)$ gauge-field, we can bound the range of ${\rm Re}(y)$. Introduce a second FI-term which couples the same field $Y$ to the new field strength $\tilde{\Sigma}$, 
\be
-\frac{{\tilde k}}{8\pi}\int d^2x d\theta^+ d\bar{\theta}^- \, Y\tilde\Sigma +\text{c.c.},
\ee
with charged fields $\tilde\phi$ satisfying 
\be
\sum_i \tilde{Q}_i |\tilde{\phi}_i|^2 = \tilde{r} - {2 \tilde{k}} {\rm Re}(y). 
\ee
As long as ${\tilde k} \leq 0$, $\tilde{r} \ge 0$, and the charges $Q_i, \tilde{Q}_i > 0$, the range of ${\rm Re}(y)$ is bounded:
\be \label{boundedrange}
\frac{{\tilde r}}{2 {\tilde k}}  \leq  {\rm Re}(Y) \leq \frac{r}{2 k}. 
\ee
This gives the cylindrical fixture of Figure~\ref{figure2}\ with a product of weighted projective spaces fibered over the cylinder. In this basic fixture, the size of each projective space vanishes at one of the ends. 

Once again, the simplest models have $n$ fields $\Phi^i$ with charges $(1,0)$ and $\tilde{n}$ fields $\tilde{\Phi}^{\tilde \imath}$ with charges $(0,1)$. Taking $k \ge 0, \tilde{k} \le 0$ as in \eqref{boundedrange}, we find the metric and $B$ field
\begin{align}
ds^2=&\ R(y)\left(\frac{dz\cdot d\bar{z}}{1+|z|^2}-\frac{|\bar{z}\cdot dz|^2}{(1+|z|^2)^2}\right)+\tilde{R}(y)\left(\frac{d\tilde{z}\cdot d\bar{\tilde{z}}}{1+|\tilde{z}|^2}-\frac{|\bar{\tilde{z}}\cdot d\tilde{z}|^2}{(1+|\tilde{z}|^2)^2}\right)\nonumber \\
&+\left(1+\frac{k^2}{R(y)}+\frac{\tilde{k}^2}{\tilde{R}(y)}\right)dyd\bar{y}, \\
B=&\  \frac{k \bar{z}\cdot dz \wedge d\bar{y}+kz\cdot d\bar{z}\wedge dy}{1+|z|^2} + \frac{\tilde{k} \bar{\tilde{z}}\cdot d\tilde{z} \wedge d\bar{y}+\tilde{k}\tilde{z}\cdot d\bar{\tilde{z}}\wedge dy}{1+|z|^2}.
\end{align}
We can calculate the $H$ flux from $B$,
\begin{align}
H=2d\theta\wedge \left(kJ_{FS}+\tilde{k}\tilde{J}_{FS}\right),
\end{align}
which is easy to integrate over the two-cycle dual to either $J_{FS}$ or $\tilde{J}_{FS}$, and the $S^1$ formed by $\theta$,
\begin{align}
\int_{\mathcal{C}\times S^1}H&=4\pi k, & \int_{\tilde{\mathcal{C}}\times S^1}H=4\pi\tilde{k}.
\end{align}
Note that while $k$ and $\tilde{k}$ must have opposite signs in order to bound $Y$, they do not necessarily have the same magnitude, and the same is true of the corresponding $H$-fluxes. This is somewhat surprising because we might have expected the total brane and anti-brane charge to sum to zero for a compact space. However, this does not seem to be a requirement for these geometries. The dual descriptions of  models  with $k \neq -\tilde{k}$, which will be discussed in section~\ref{sect:doubletrumpetkahler},  involve either squashed weighted projective spaces, or spaces with orbifold singularities.

When each fibered projective space is actually a sphere, which happens for $\PP^1 \sim S^2$, the resulting space is $S^5 \times S^1$~\cite{Quigley:2012gq}. Otherwise the space looks singular and, as we will discuss shortly, we will need the T-dual description to see that the collapsing projective space is actually acceptable.

Near each end, the metric has the same asymptotic form as \eqref{metriconeu1y}, with the extra projective space staying at finite size. The spaces discussed in this section so far have been previously studied in a $(0,2)$ context in \cite{Quigley:2012gq}.

\subsubsection{Compact and conformal models?} \label{compactconformal}

The model with two $U(1)$ gauge fields gives us a construction of a torsional compact geometry. However, there is no $U(1)_R$ charge assignment for $y$ that allows us to solve \eqref{eq:conformalityconditionnop}, and so this is a massive model. We can show that compact models will be generically massive if we only allow $\Phi$ and $Y$ fields with couplings of the form $Y\Sigma$ and usual chiral superpotentials.

As we have seen in the previous sections, we obtain one bound on the range of $\rept(y)$ from each $D$-term condition, as long as all fields $\Phi^i$ charged under the corresponding $U(1)$ have positive charges (or all negative charges). The bound is of the form
\begin{align}
\frac{\sum_i Q_i |\phi^i|^2}{\sum_j Q_j} \ge 0 \Rightarrow \frac{r-2k\rept(y)}{\sum_j Q_j} \ge 0 \Rightarrow \frac{2k\rept(y)}{\sum_j Q_j} \le \frac{r}{\sum_j Q_j}.
\end{align}
This allows us to see that the direction of this bound depends only on the sign of $\frac{\sum_j Q_j}{k}$. But that sign is precisely what sets the sign of $\gamma$ satisfying \eqref{eq:conformalityconditionnop}. Therefore if we have two bounds in opposite directions as needed to make the range of $\rept(y)$ compact, there is no $\gamma$ which solves \eqref{eq:conformalityconditionnop} for both $U(1)$, and we will be dealing with a massive model.

We may now illustrate this argument with an example of an attempt to evade it, in order to provide some intuition on how compactness is violated. Since the effects of $Y$ are generally not enough to cancel the $U(1)_R$ anomalies from two $U(1)$ gauge fields, we can try to also add a negatively-charged field coupled in a superpotential. This helps with cancelling the anomaly, as it usually does for example in the quintic. However, since the models we have been considering have a point where the size of the ambient projective space vanishes, we would be left with a non-compact direction where $Y$ and the negatively-charged field grow without bound. We can try to be smarter and add an extra $U(1)$.

Building on the double trumpet model, take then three $U(1)$ gauge fields and the following field content: $n$ fields $\Phi^i$ with charges $(1,0,0)$, $\tilde{n}$ fields $\tilde{\Phi}^i$ with charges $(0,1,0)$, a field $S$ with charges $(-Q,-\tilde{Q},0)$ and a field $A$ with charges $(0,0,Q_a)$. Introduce a periodic twisted chiral $Y$, coupled to the three twisted chirals $\Sigma, \tilde{\Sigma}, \Sigma_a$ with coefficients $(k, \tilde{k}, k_a)$. As before we will want $k \tilde{k}<0$ in order to obtain a bound, so choose $\tilde{k} < 0$. The field $S$ can be used to write a gauge-invariant superpotential $S f(\Phi) g(\tilde{\Phi})$, with $f$ and $g$ polynomials of degrees $Q$ and $\tilde{Q}$, respectively. The $D$-term constraints read
\begin{subequations}
\begin{align}
-Q|s|^2+|\phi|^2&=r-2k\rept(y), \label{eq:Dtermcompconf1} \\
-\tilde{Q}|s|^2+|\tilde{\phi}|^2&=\tilde{r}-2\tilde{k}\rept(y), \label{eq:Dtermcompconf2} \\
Q_a|a|^2&=r_a-2k_a\rept(y). \label{eq:Dtermcompconf3}
\end{align}
\end{subequations}
From this it is clear that at the previous extrema of $\rept(y)$, $\phi$ or $\tilde{\phi}$ become zero, liberating $s$ and $y$ in a non-compact direction. One of these can be removed by setting $Q=0$. Once we have done that, our anomaly cancellation conditions \eqref{eq:conformalityconditionnop} read
\begin{subequations}
\begin{align}
n-\gamma k&=0, \label{eq:conformalityincompact1} \\
\tilde{n}-\tilde{Q}-\gamma\tilde{k} &=0, \label{eq:conformalityincompact2} \\
Q_a-\gamma k_a&=0.\label{eq:conformalityincompact3}
\end{align}
\end{subequations}
Now we would like to use the bound from the third $U(1)$ to remove the second non-compact direction from the region $\tilde{\phi}=0$. However, note~\eqref{eq:conformalityincompact1}\ sets $\gamma k = n > 0 \Rightarrow \gamma >0$. Using \eqref{eq:conformalityincompact3}, this implies $Q_a/k_a >0$. Then the constraint imposed on $y$ from~\eqref{eq:Dtermcompconf3}\ will have the same sign as that from~\eqref{eq:Dtermcompconf1}, and therefore will not help when the second D-term constraint disappears and $S$ becomes non-zero.

\subsection{The K\"ahler picture} \label{sect:kahlerpicture}
Both examples in sections \ref{sect:ysigmaoneu1} and \ref{sect:ysigmatwou1} have T-duals which are K\"ahler and described by the squashed toric varieties first discussed in~\cite{Hori:2001ax}. At the level of an ultraviolet GLSM, squashing is implemented as follows. 

Consider a toric GLSM, namely a collection of $n$ chiral superfields charged under a collection of $k$ abelian gauge symmetries. Such a model, in the absence of a superpotential, has $n-k$ remaining flavor symmetries. The squashing construction gauges each of these flavor symmetries while simultaneously adding a Stueckelberg chiral superfield for each. An action for such a model is
\begin{align}\label{eq:generalsquashedtoric}
S &= \frac{1}{16\pi} \int d^2x d^4 \theta \left( \sum_{i = 1}^n\sum_{a=1}^k\sum_{\alpha=1}^{n-k} |\Phi_i|^2e^{2Q_i^aV_a+2R_i^\alpha V_\alpha} - \sum_{a = 1}^k \frac{2\pi}{e_a^2}|\Sigma_a|^2 \right) \nonumber \\
&-\frac{i}{8\pi} \int d^2x d\theta^+d\bar{\theta}^- \sum_{a=1}^k t^a \Sigma_a + \text{c.c} \\
& + \frac{1}{16 \pi} \int d^2x d^4\theta \sum_{\alpha = 1}^{n-k}\left( \frac{b_\alpha}{2}\left( P_\alpha+\bar P_\alpha + 2 V_\alpha\right)^2 -\frac{2\pi}{e_\alpha^2} |\Sigma_\alpha|^2 \right). \nonumber
\end{align}
The charges of the chiral fields under the original $k$ gauge symmetries are $Q_i^a$, while the charges of the flavor symmetries are $R_i^\alpha$. We stipulate that the combined $n\times n$ matrix $(Q_i^a, R_i^\alpha)$ has rank $n$. Note also that there are no FI couplings for the gauged flavor symmetries, as these can be absorbed into a redefinition of the corresponding Stueckelberg fields.

Each squashing has an associated squashing parameter $b_\alpha \in \mathbb{R}$. In the limit $b_\alpha \to \infty$, the Stueckelberg fields decouple, and the squashing is removed. Since each squashing corresponds to a $U(1)$ isometry, the Stueckelberg fields are periodic: $\mathrm{Im}P_\alpha \sim \mathrm{Im}P_\alpha + 2\pi$. Modifying the charge of $P_\alpha$ to ${k_\alpha}$ for some $k_\alpha\in \mathbb{Z}$ yields a $\mathbb{Z}_{k_\alpha}$ orbifold~\cite{Hori:2001ax}. Note that in section~\ref{sect:boundingdterms} we took $b_\alpha=1$ for all fields $Y$. Taking the limit of no squashing in the $Y$-picture leads to a $Y$ field that has a vanishing classical kinetic term but gains a metric when we descend to the non-linear sigma model. The torsion of the $Y$ model is unaffected by the value of the squashing parameter.

Applying the T-duality of section~\ref{sect:Tduality} results in the field-dependent FI couplings we have already seen. The T-dual of \eqref{eq:generalsquashedtoric} is
\begin{align}
S &= \frac{1}{16\pi} \int d^2x d^4 \theta \left( \sum_{i = 1}^N\sum_{a=1}^k\sum_{\alpha=1}^{N-k} |\Phi_i|^2e^{2Q_i^aV_a+2R_i^\alpha V_\alpha} -  \sum_{a = 1}^k \frac{2\pi}{e_a^2}|\Sigma_a|^2 \right) \nonumber \\
& -\frac{i}{8\pi} \int d^2x d\theta^+ d\bar{\theta}^- \sum_{a=1}^k t^a \Sigma_a + \text{c.c} \\
& - \frac{1}{16 \pi} \int d^2x d^4\theta \sum_{\alpha = 1}^{N-k} \left( \frac{\bar{Y}_\alpha Y_\alpha}{b_\alpha} + \frac{2\pi}{e_\alpha^2} |\Sigma_\alpha|^2 \right)\nonumber \\
&- \frac{1}{8\pi} \int d^2x d\theta^+d\bar{\theta}^- \sum_{\alpha=1}^{N-k}Y_\alpha \Sigma_\alpha + \text{c.c}. \nonumber
\end{align}
Dualizing a field $P_\alpha$ with shift-charge $k_\alpha$ would lead to a factor $k_\alpha$ multiplying the last line, just like the couplings $k$ in section~\ref{sect:boundingdterms}. We will also mostly take $b_\alpha=1$ to connect to the previous discussion.


\subsubsection{One \texorpdfstring{$U(1)$}{U(1)} action}
The dual of the trumpet geometry discussed in section~\ref{sect:ysigmaoneu1} has $n$ chiral fields of charge $1$ and a Stueckelberg field of charge $k$ and period $2\pi$ with kinetic action
\begin{equation*}
\frac{1}{32 \pi} \int d^2x d^4\theta~ \left( P + \bar P + 2k V\right)^2.
\end{equation*}
This model was discussed in~\cite{Hori:2002cd}. The total space is topologically $\mathbb{C}^n/\mathbb{Z}_k$, but it is to easier to visualize as $S^{2n-1}$ warped over a half line $R_+$. The orbifold action is through discrete translations along the fiber of the Hopf fibration $U(1) \hookrightarrow S^{2n-1} \rightarrow \mathbb{C}P^n$. Note that if $k=1$, this space is birationally equivalent to the total space of the tautological bundle over $\mathbb{C}P^{n-1}$. The classical metric in an affine patch of the $\mathbb{C}P^{n-1}$ base is
\begin{equation}
ds^2 = \rho^2\left(\frac{dz\cdot d\bar{z}}{1+|z|^2}-\frac{|\bar{z}\cdot dz|^2}{(1+|z|^2)^2}\right)+\left(1+\frac{\rho^2}{k^2}\right)d\rho^2+\frac{\rho^2}{\left(1+\frac{\rho^2}{k^2}\right)} \left( \frac{d\theta}{k} + A_{FS} \right)^2,
\end{equation}
where $\theta$ has period $2\pi$ and $A_{FS}$ in this patch is
\begin{equation}
A_{FS} = -\frac{i}{2} \frac{z\cdot d\bar z - \bar z \cdot dz}{1 + |z|^2}.
\end{equation}
This is, indeed, the T-dual of \eqref{metriconeu1y} using the $B$-field \eqref{bfieldoneu1y}.

This type of T-duality also provides us with a recipe for T-dualizing in the UV GLSM along a given circle isometry in a non-linear sigma model. Once we identify the GLSM circle that descends to the circle we would like to dualize, we can gauge the flavor symmetry corresponding to that isometry, and add a corresponding Stueckelberg field $P$. We can then dualize $P$ into a $Y$ as in Appendix~\ref{sect:Tdualityexpanded}, and descend to the non-linear sigma model in the resulting theory. Finally, if we want to remove the effects of the new field, we should take the limit of no squashing, or $b \rightarrow \infty$. It would be interesting to try this prescription in models with blowing-up circles, such as those of \cite{Melnikov:2012nm}.

\subsubsection{Two \texorpdfstring{$U(1)$}{U(1)} actions} \label{sect:doubletrumpetkahler}
This dual has $n$ chirals of charge $(1,0)$ and $\tilde{n}$ chirals of charge $(0,1)$ and a single Stueckelberg with charge $(k,\tilde k)$ and period $2\pi$ with kinetic term
\begin{equation*}
\frac{1}{32 \pi} \int d^2x d^4\theta~ \left( P + \bar P + 2k V + 2\tilde k \tilde V\right)^2.
\end{equation*}
As before, we will assume that $k>0$ and $\tilde k <0$ in order that the target space be compact. Let $\h{k} = \gcd(k,-\tilde k)$. We can perform an $SL(2,\mathbb{Z})$ transformation on the gauge fields
\begin{equation}
\begin{pmatrix} \h{V} \\ \check{V} \end{pmatrix} = \begin{pmatrix} \frac{k}{\h{k}} & \frac{\tilde k}{\h{k}} \\ -\tilde \ell &  \ell \end{pmatrix}\begin{pmatrix} V \\ \tilde V \end{pmatrix}, \qquad k \ell + \tilde k \tilde \ell = \h{k},
\end{equation}
yielding a theory with $n$ chirals of charge $(\ell, -\frac{\tilde k}{\h{k}})$ and $\tilde n$ chirals of charge $(\tilde \ell, \frac{k}{\h{k}})$ and a Stueckelberg field of charge $(\h{k}, 0)$ and period $2\pi$ with kinetic term
\begin{equation*}
\frac{1}{32 \pi} \int d^2x d^4\theta~ \left( P + \bar P + 2\h{k} \h{V}\right)^2.
\end{equation*}
Matching onto the general model, we see this describes a weighted $\mathbb{C}P^{n + \tilde n -1}$ with $n$ weights of $-\frac{\tilde k}{\h{ k}}$ and $\tilde n$ weights of $\frac{k}{\h{k}}$. Recall that all these weights are positive because $\tilde{k}<0$. Further, this space is squashed and orbifolded along the $U(1)$ isometry under which the $n$ chirals have charge $\ell$ and the $\tilde n$ chirals have charge $\tilde \ell$.

\subsection{More general fibrations}\label{generalcase}


The models described so far have both a K\"ahler and a torsional description, related by duality. We would like to find models which do not have a K\"ahler description, and hence live beyond the lamp post.  

There is a very natural way to construct such models. Imagine a base GLSM theory with $Y$-fields and some charged $\Phi$ fields. We will fiber a twisted sigma model over this base theory in a way that obstructs dualizing $Y$ back to a chiral superfield. We will fiber the complex structure of the twisted sigma model over $Y$ using superpotential couplings between $Y$ and charged twisted chirals $\h{\Phi}$. 

As a first example modeled on the quintic, consider a theory with a $U(1)^2$ gauge group with charged chirals and a $\h{U}(1)$ gauge group with charged twisted chirals. The charge matrix is given in Table~\ref{tab:generalfibration}.
\begin{table}
    \centering
    \begin{tabular}{l c c c}\toprule
& $U(1)_1$ & $U(1)_2$ & $\h{U}(1)$\\ \midrule
$\Phi$ & 1 & 0 & 0\\
$\tilde{\Phi}$ & 0 & 1 & 0 \\
$\h{\Phi}$ & 0 & 0 & 1 \\
$\h{S}$ & 0 & 0 & $-5$ \\ \bottomrule
\end{tabular}
    \caption{\textit{Charge matrix for the quintic fibration over the double trumpet.}}
    \label{tab:generalfibration}
\end{table}



We will not worry about imposing conformal invariance for the moment. Rather, our interest is in the structure of the resulting generalized K\"ahler geometries. The chiral content is the same as in section~\ref{sect:ysigmatwou1}, and so the range of $\rept(y)$ will be bounded. We do not need to assign any $R$-symmetry transformation to $Y$. On the other hand, the twisted chiral content is similar to the usual quintic, except that we allow the twisted superpotential to depend on $e^Y$:
\begin{equation}
\h{W}=\h{S}f(e^Y, \h{\Phi}),
\end{equation}
where $f$ is a polynomial of degree 5 in $\h{\Phi}$ whose coefficients are polynomials in $e^Y$.

For concreteness, we can consider deforming the Fermat quintic by a $Y$-dependent monomial, so
\begin{equation}
    f(e^Y, \h{\Phi}) = \sum_{i=1}^5 \h{\Phi}_i^5 + e^Y \h{\Phi}_1 \h{\Phi}_2 \h{\Phi}_3 \h{\Phi}_4 \h{\Phi}_5.
\end{equation}
This $Y$-dependence cannot be removed by a field redefinition; the complex structure modulus parametrized by $\h{\Phi}_1 \h{\Phi}_2 \h{\Phi}_3 \h{\Phi}_4 \h{\Phi}_5$ is now fibered over the $Y$ cylinder. This space is therefore a fibration of the quintic CY $3$-fold over the double trumpet model. For a suitable choice of parameters, the complex structure of the fiber can be kept away from degeneration limits. This is the structure illustrated in Figure~\ref{figure3}.

It is easy to generalize this structure to more general fibrations and bases resulting in compact intrinsically torsional spaces. If, however, one wishes to impose conformal invariance on the resulting space then we encounter the flat direction issue described in section~\ref{compactconformal}. The other natural possibility is to include $e^Y$ couplings to ${\S}$ fields. That case is rather interesting and will be discussed in section~\ref{sect:expysigma}.

\subsubsection{A comment on squashed Calabi-Yau}

Although we run into the flat direction issue when trying to build compact conformal models from $Y$ fixtures, we can certainly 
repeat the usual hypersurface or complete intersection construction of compact Calabi-Yau spaces for squashed projective spaces in terms of $P$ variables. 

Let us describe these models by way of an example: a squashed analogue of the quintic Calabi-Yau three-fold. Specifically, the model will describe a hypersurface inside of a squashed $\mathbb{C}P^4$ that is topologically Calabi-Yau. Recall the squashed $\mathbb{C}P^4$: start with a model with $5$ chiral fields $\Phi_i$, each with charge $1$ under a single $U(1)_G$ gauge symmetry. We choose to squash the flavor symmetry $U(1)_F$ under which only $\Phi_5$ rotates with charge 1. To do this, we gauge this symmetry and add a chiral Stueckelberg field $P$ with kinetic term
\begin{equation*}
\frac{b}{32\pi} \int d^4\theta~ \left( P + \bar P + 2 V_F \right)^2,
\end{equation*}
where $V_F$ is the vector superfield of the gauged flavor symmetry. As usual, we choose $\mathrm{Im} P$ to have period $2\pi$. In order that the squashed $\mathbb{C}P^4$ be smooth, we choose the shift charge of $P$ to equal one; we can effect an orbifold of the ambient space by choosing another integer charge $\ell$.

To carve out a hypersurface, we include another chiral field $S$ with charge $-5$ under the original gauge symmetry and charge $0$ under the flavor symmetry, and we add the superpotential
\begin{equation*}
W(S,\Phi_i,e^P) = S \left( G(\Phi_1, \ldots, \Phi_4) + e^{-5P} \Phi_5^5 \right).
\end{equation*}
The polynomial $G$ is homogeneous of degree $5$ and non-singular in the subspace $\Phi_5 = 0$. A summary of the fields and their charges is provided in Table~\ref{tab:squashedquintic}.

\begin{table}
\begin{center}
\begin{tabular}{l c c }
\toprule
& $U(1)_G$ & $U(1)_F$ \\ \midrule
$\Phi_{1-4}$ & 1 & 0 \\
$\Phi_5$ & 1 & 1 \\
$S$ & $-5$ & 0 \\
$P$ & 0 & 1 \\ \bottomrule
\end{tabular}
\caption{\textit{Charge matrix for the squashed quintic model.}}
\label{tab:squashedquintic}
\end{center}
\end{table}

This model is of the form described in \eqref{eq:generalPYmodel}, and we can build a protected superconformal multiplet in $\bar Q_+$-cohomology with the following parameter choices
\begin{equation}
\gamma = -\beta = 1, \qquad \alpha_5 = -1, \qquad \alpha_i = 0,\, \ i = 1, \ldots, 4, \qquad \alpha_S = 1.
\end{equation}
As usual, this choice is ambiguous up to shifts by the charges under the original $U(1)$ gauge action. Now we expect this construction to give the correct central charge for the non-compact total space of $O(-5)$ over the squashed $\mathbb{C}P^4$, prior to turning on the superpotential:
\begin{equation}
c_{\rm non-cpt} = 9 + \frac{6}{b}.
\end{equation}
This central charge depends on the squashing parameter $b > 0$. However, there is an interesting question of the correct IR description of this theory with the superpotential turned on. By a field redefinition $\phi_5 \rightarrow e^{-p}\phi_5$, the effect of the squashing can be removed from the F-term constraints. The only effect of the squashing is a change of D-terms. However, the IR theory is expected to be fully determined by the F-term structure if the metric is compact, so our expectation is that this theory flows to the usual quintic Calabi-Yau conformal field theory with $c=9$ in the IR.\footnote{We would like to thank Ilarion Melnikov for clarifying this issue.} 

What we do learn from this construction is that there should be a corresponding relevant operator in $Y$ variables that also produces a compact CFT. The $Y$ description contains two $U(1)$ gauge-fields with $D$-term constraints, 
\bea
|\phi_1|^2 +|\phi_2|^2+|\phi_3|^2+|\phi_4|^2+|\phi_5|^2 - 5|S|^2 = r_1, \\
|\phi_5|^2 = r_2 - 2 {\rm Re}(y). 
\eea
A part of that relevant operator is just the superpotential couplings involving fields that are not dualized:
\begin{equation*}
W(S,\Phi_i,e^P) = S\left(G(\Phi_1, \ldots, \Phi_4) \right).
\end{equation*}
This interaction still leaves flat directions for the physical potential. To lift those remaining flat directions, we need the dual of the chiral operator $e^{-5P} \Phi_5^5$. However, as is often found in dual descriptions, this nice local chiral operator has no simple local description in  $Y$ variables. 

\subsection{Exponential couplings} \label{sect:expysigma}

We now consider more exotic models that also obstruct a straightforward dualization to a K\"ahler picture. These models break the $U(1)$ symmetry shifting ${\rm Im}(Y)$. That circle is precisely the one we would want to T-dualize to produce a K\"ahler picture in terms of $P$ variables.

\subsubsection{One \texorpdfstring{$U(1)$}{U(1)} action}\label{sect:expysigmaoneu1}

As a first example of such a model, consider a theory with one $Y$, one $U(1)$ gauge-field and a twisted superpotential
\begin{align}
S = - \frac{1}{8\pi} \int d^2x d\theta^+ d\bar{\theta}^- \kappa e^Y \Sigma.
\end{align}
This model is the analogue of Figure~\ref{fg}, and the new coupling expressed  in component fields is found in~\eqref{eqn:expysigmacomponents}. This  coupling explicitly breaks the $U(1)$ isometry which shifts the imaginary part of $Y$. It should also be noted that while the coupling $k$ in previous sections was integer, $\kappa$ can take any real value. This coupling also fixes $Y$ to be invariant under $R$-symmetry transformations, so $Y$ can no longer be used to absorb any possible anomalies. 

\begin{figure}[ht!]
\centering
\includegraphics[scale=0.55]{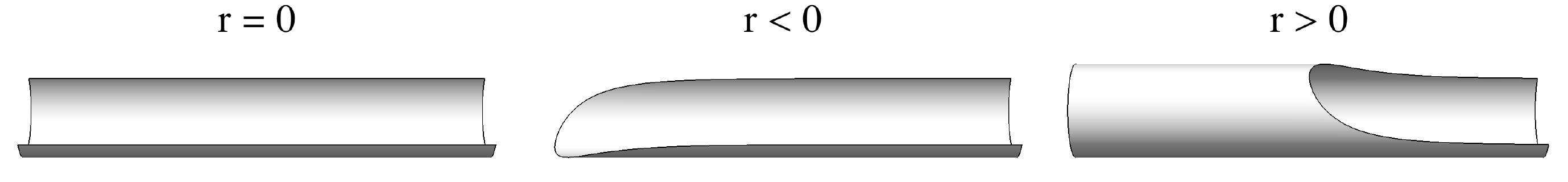}
\caption{\textit{The three possibilities.}}\label{figure4}
\end{figure}

Writing $y=a+i\theta$, the $D$-term potential condition now reads
\begin{equation}
\sum_i Q_i |\phi_i|^2=r-2\kappa e^a\cos \theta=R(y),
\end{equation}
so if all the charges $Q_i$ are positive the space is bounded to the region
\begin{equation} \label{rbound}
2\kappa e^a\cos \theta \leq r.
\end{equation}
Topologically, the space of possible $y$ values satisfying this inequality can have one of three shapes, depending only on $r$. These three possibilities are depicted in Figure~\ref{figure4}. 
\begin{itemize}
\item
If $r=0$, the condition~\C{rbound}\ picks out one sign of the cosine for any $a$, and we obtain an infinite strip. 

\item
If $r < 0$, the condition imposes a bound on $a$, and we find a semi-infinite strip. 

\item 
If $r > 0$, on the other hand, we essentially obtain the converse of that, an infinite cylinder with a semi-infinite strip taken out.
\end{itemize}

The metric and $B$-field can be obtained using similar methods from the above examples, and have the form
\begin{align}
ds^2=&\ R(y)\left(\frac{dz\cdot d\bar{z}}{1+|z|^2}-\frac{|\bar{z}\cdot dz|^2}{(1+|z|^2)^2}\right)+\left(1+\frac{\kappa^2e^{2a}}{R(y)}\right)dyd\bar{y}, \\
B=&\  \frac{\kappa e^{\bar{y}} \bar{z}\cdot dz \wedge d\bar{y}+\kappa e^yz\cdot d\bar{z}\wedge dy}{1+|z|^2}. \label{eqn:Bfieldexpysigma}
\end{align}
As above, the metric blows up at the boundary $R(y)=0$. Define
\begin{align}
\rho^2 &= \tilde{r}-2\kappa e^a \cos \theta, \\
\alpha &= -\kappa e^a \sin \theta,
\end{align}
so
\begin{equation}
dyd\bar{y}= da^2+d\theta^2=\frac{\rho^2 d\rho^2+d\alpha^2}{\kappa^2 e^{2a}},
\end{equation}
and the metric becomes, in the limit close to the boundary,
\begin{equation}
ds^2= R(y)ds^2_{FS}+\rho^2\tilde{ds}^2_{FS}+d\rho^2+\frac{1}{\rho^2}d\alpha^2,
\end{equation}
which has a similar form to the metric close to the boundaries in trumpet models, as can be seen by comparison with~\eqref{metriconeu1y}.

The torsion in this case can be obtained from \eqref{eqn:Bfieldexpysigma} and has the form
\begin{equation}
H=2\kappa d(e^a\sin\theta)\wedge J_{FS}=-2d\alpha \wedge J_{FS}
\end{equation}
This three-form integrates to zero in any region where the $\theta$ circle closes, but has a non-zero integral in other regions.

\subsubsection{Several \texorpdfstring{$U(1)$}{U(1)} actions} \label{sect:expysigmaseveralu1}

In order to build a compact space including an exponential coupling to a field strength multiplet, start by taking the field content of section~\ref{sect:ysigmatwou1}\ which led to Figure~\ref{figure2}. This theory includes two $U(1)$ gauge fields $\Sigma$ and $\tilde{\Sigma}$ coupled to two sets of chiral fields $\Phi^i$ and $\tilde{\Phi}^i$, leading to the D-term conditions:
\begin{align}
\sum_i Q_i |\phi^i|^2 &= r - {2 k} {\rm Re}(y) = R(y), \\
\sum_i \tilde{Q}_i |\tilde{\phi}^i|^2 &= \tilde{r} - {2 \tilde{k}} {\rm Re}(y) = \tilde{R}(y).
\end{align}
This bounds the range of $\rept(y)$ if all charges $Q_i$ and $\tilde{Q}_i$ are positive, $k>0$, and $\tilde{k}<0$. To this configuration, which is dual to a squashed space,  we can now add a third $U(1)$ multiplet $\Sigma'$ with its own set of charged fields $\Phi'$, coupled to $Y$ with a superpotential $\kappa e^Y \Sigma'$. The corresponding D-term condition reads
\begin{equation}
    \sum_i Q_i' |\phi_i'|^2=r'-2\kappa e^a\cos \theta = R'(y).
\end{equation}
The boundaries of the space are set by the $y$ values where any single projective space collapses to zero size. We encounter a problem with new flat directions if any two projective spaces collapse to zero size at the same $y$ value. To see this, note that the $Y$ superpotential can only mass up a single combination of $\S$ fields. If two or more projective spaces collapse at the same point, there will be two distinct $U(1)$ factors for which no charged fields have an expectation value at that point. Therefore, one $\S$ multiplet will be massless resulting in a new flat direction.


It is straightforward to see that to avoid an intersection where two or more projective spaces collapse, we need to be in the case where the exponential allows the full $y$ circle, so $r'>0$. This is the last case depicted in Figure~\ref{figure4}. We also need the boundary for the projective space with the exponential coupling to be fully outside the space defined by the other two constraints. The set of $a$ satisfying the condition
\begin{equation}
e^a\ge\frac{r'}{2|\kappa|},
\end{equation}
contains a boundary point for some $\theta$ where the projective space with the exponential coupling vanishes. We therefore need to impose the condition 
\begin{equation}\label{eqn:expysigmanointersection}
\log\frac{r'}{2|\kappa|}>\frac{r}{2k}, 
\end{equation}
to ensure these boundary points are excluded. We will define the theory at a UV scale $\Lambda$ with bare FI parameters satisfying inequality~\eqref{eqn:expysigmanointersection}. The inequality will then generally be preserved by RG flow because the additive renormalization of the FI parameters makes the right-hand side decrease faster than the left-hand side as we flow down in energy.

\begin{figure}[ht!]
\centering
\includegraphics[scale=0.45]{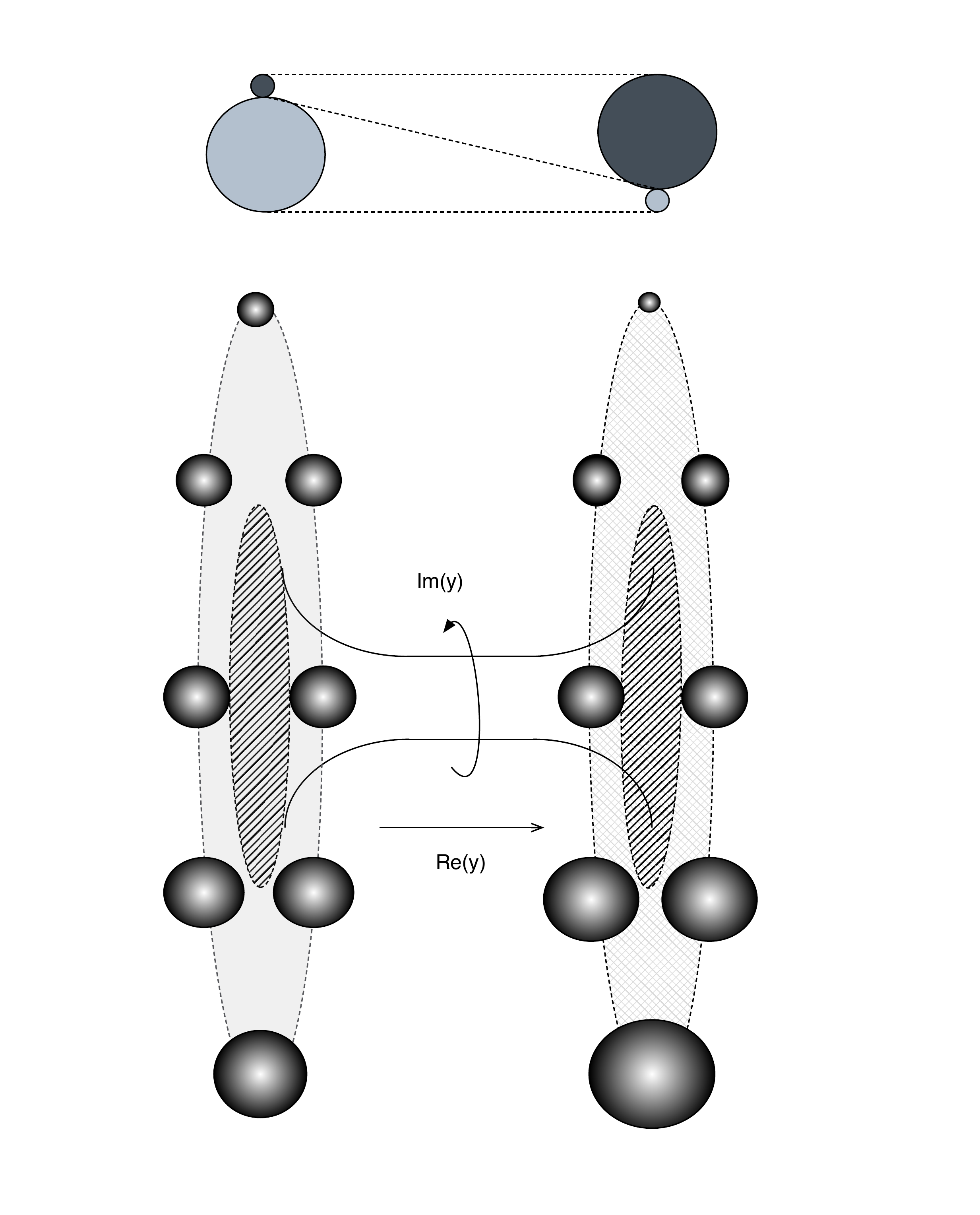}
\caption{\textit{A compact example with exponential couplings. The three fibrations are depicted. The first two fibrations are the ones already seen in Figure~\ref{figure2}. The radius of the third space depends on both the real and imaginary parts of $y$. Note that the radius oscillations become larger as $\rept(y)$ becomes bigger.}}\label{figure5}
\end{figure}

The metric and $B$-field are essentially the sum of those given in sections~\ref{sect:ysigmatwou1} and \ref{sect:expysigmaoneu1}. This model then consists of the same ingredients as those of the double trumpet model, with an additional space fibered over the double trumpet whose size depends on the real and imaginary values of $y$. This fibration structure is depicted in Figure~\ref{figure5}.  The integrated torsion will have the same value as found in the double trumpet. However, since the isometry is broken by the exponential coupling, this model cannot be dualized into a K\"ahler picture like a squashed projective space.

\subsection{Unifying constructions} \label{sect:fysigma}

We can unify the structures described in sections \ref{sect:boundingdterms} and \ref{sect:expysigma} by writing the $\S^a$ coupling as $f_a(Y)\Sigma^a$, with an $f_a(Y)$ that shifts at most by an integer multiple of $2\pi i$ when $Y$ shifts by $2\pi i$. For simplicity, take each $\phi$ to have charge 1 under one of the gauge symmetries, and 0 under the others. Then the $D$-terms constrain
\begin{align}
\sum_i|\Phi_i^a|^2 e^{2A_a}=r_a-2\rept f_a(y)=R_a(y),
\end{align}
and repeating the analysis yields a metric of the form
\begin{align}
ds^2=\sum_a R_a(y)ds^2(\phi^a)+\sum_\mu |dy_\mu|^2+\sum_a \frac{|df_a|^2}{R_a(y)},
\end{align}
Near $R_a=0$ it makes sense to choose coordinates including $\rho^2=f_a$, and in that limit the $\rho$ metric will reduce to the familiar form~\eqref{metriconeu1y}. The $B$ field can also be written in terms of the functions $f_a$ as
\begin{align}
B&=\sum_a \frac{\phi^a\cdot d\bar{\phi}_a \wedge df_a}{1+|\phi^a|^2}+\text{c.c.} \\
\Rightarrow H&=\sum_a 2d\left(\impt f_a\right) \wedge J^a_{FS}.
\end{align}
If the circle from the imaginary part of $y$ closes, we can see from this expression that the integral of $H$ will be non-zero if $f_a$ is not single-valued when we travel around the circle.

\section{The Quantum Cohomology Ring}\label{sect:quantumcoh}

Our discussion so far has been largely classical. We wanted to describe gauge theories that give rise to classical vacuum equations describing generalized K\"ahler spaces. In this section, we turn to some quantum aspects of these models. Specifically, we will probe the vacuum structure by calculating the quantum cohomology rings for some of the models we have discussed.

\subsection{Coulomb branch vacua} \label{sect:coulombvacuaderiv}

In $(2,2)$ gauge theories, we can investigate vacua where the scalars in field strength multiplets $\Sigma_a$ gain expectation values. In this section, we review the basic ingredients needed to describe these Coulomb branch vacua. We can see from~\eqref{chiralkinetic} that all fields $\phi^i$ charged under the gauge symmetries become massive when each $\S_a$ has an expectation value. Such fields can then be integrated out at one loop, leading to a quantum correction to the twisted superpotential. The effective superpotential has the form \cite{Witten:1993yc,Morrison:1994fr,Melnikov:2005hq}
\begin{align}
S_{\widetilde{W}}&=\frac{1}{8\pi}\int d^2x d\theta^+ d\bar{\theta}^- \sum_a \Sigma_a\left(\sum_i Q_i^a\left[\log\left(\frac{\sum_b Q_i^b\Sigma_b}{\Lambda}\right)-1\right]-it_a\right) \nonumber \\
&=\frac{1}{8\pi}\int d^2x d\theta^+ d\bar{\theta}^- \sum_a \Sigma_a\left(\sum_i Q_i^a\left[\log\left(\frac{\sum_b Q_i^b\Sigma_b}{\mu}\right)-1\right]-it_a(\mu)\right)
\end{align}
where $\Lambda$ is the UV renormalization scale, $t_a$ are the bare FI parameters, and $t_a(\mu)$ include the renormalization
\begin{equation} \label{eqn:firunning}
t_a(\mu)=t_a(\Lambda)+i\left(\sum_i Q_i^a\right)\log\frac{\mu}{\Lambda}.
\end{equation}
Note that fields $P$ that are shift-charged under the gauge symmetries or fields $Y$ with a $Y\Sigma$ superpotential are not massed up in the same way. This expression is only valid at large values of $\Sigma_a$, where the masses of the fields that have been integrated out are large, so once a solution is obtained it should be checked that we are in the right regime. Varying this superpotential leads to the vacuum equations
\begin{equation}
\sum_i Q_i^a\log\left(\frac{\sum_b Q_i^b\Sigma_b}{\mu}\right)=it_a(\mu) \Rightarrow \prod_i \left(\frac{\sum_b Q_i^b\Sigma_b}{\mu}\right)^{Q_i^a} = e^{it_a(\mu)}.
\end{equation}
Taking the exponential of both sides here does not alter the solutions to the equation, since the right-hand side includes $i\theta$, with its $2\pi i$ periodicity. Since we will be discussing generalizations of this structure, we note that supersymmetric vacua on the Coulomb branch are determined by solutions to 
\be
\exp\left({ \frac{\partial {\widetilde W}_{\rm eff}}{\partial \S_a}}\right) = 1 
\ee
for all field strength multiplets $\Sigma_a$, where ${\widetilde W}_{\rm eff}$ is the effective twisted chiral superpotential obtained by integrating out all the charged fields. If there are additional twisted chiral fields which are not field strength multiplets, like $Y$ fields, then we also impose the condition
\be
{ \frac{\partial {\widetilde W}_{\rm eff}}{\partial Y}} = 0  
\ee
for each such field $Y$. 

As a warm-up for the models considered in this work, we can use this superpotential to find the vacuum structure of the $\mathbb{C}P^{n-1}$ gauged linear sigma model. The field content of this model consists of one $\Sigma$ and $n$ chiral fields $\Phi^i$ of charge 1. A straightforward application of the formulae above leads to the superpotential
\begin{equation}
S_{\widetilde{W}}=\frac{1}{8\pi}\int d^2x d\theta^+ d\bar{\theta}^- \Sigma\left(n\left[\log\left(\frac{\Sigma}{\mu}\right)-1\right]-it(\mu)\right),
\end{equation}
giving $n$ critical points obeying
\begin{equation}
\Sigma^n=\mu^n e^{it(\mu)}. \label{eqn:cpncoulombvacuum}
\end{equation}
As we move to lower energies, we can see from~\eqref{eqn:firunning} that $\rept(it(\mu))=-r$ grows large. This means that the mass of the fields $\Phi^i$ that were integrated out, given by $|\Sigma|$, grows relative to $\mu$, and therefore the calculation is justified.

\subsection{Double trumpet} \label{sect:doubletrumpetring}

We now analyze the simplest model we presented with a bounded range for $y$, the double trumpet model introduced in section~\ref{sect:ysigmatwou1}. To study the Coulomb branch, we take $\Sigma, \tilde{\Sigma}$ to have non-zero vacuum expectation values. This causes the fields $\Phi, \tilde{\Phi}$ to become massive, and we can integrate them out. Note that $Y$ is not massed up by these expectation values, since there is no $y\sigma$ potential. The effective superpotential was computed in section~\ref{sect:coulombvacuaderiv}.
The equations determining supersymmetric vacua take the form:
\begin{subequations}
\begin{align}
k \Sigma+\tilde{k}\tilde{\Sigma}&=0, \label{eqn:ysigmacoulomb1} \\
n\log\left(\frac{\Sigma}{\mu}\right)- k Y&=i t, \label{eqn:ysigmacoulomb2} \\
\tilde{n} \log \left(\frac{\tilde{\Sigma}}{\mu}\right) - \tilde{k} Y &= i \tilde{t}, \label{eqn:ysigmacoulomb3}
\end{align}
\end{subequations}
These three equations can be solved for $Y,\Sigma$ and $\tilde{\Sigma}$, leading to a set of ring relations between these twisted chiral operators. 

There are some basic issues to understand in this $Y$ picture. The first issue is one of identifying observables. In conventional K\"ahler GLSM models, each $\S_a$ field is associated to an FI parameter and therefore to a K\"ahler class. For non-K\"ahler models, the map between Coulomb branch operators and observables on the Higgs branch is not a priori clear. For this particular model, each field $\S, \tilde{\S}$ and $Y$ is a twisted chiral superfield with a bottom component that is $(\bar{Q}_++Q_-)$-closed, but not exact. The $Y$ field is distinguished from the $\S,\ \tilde{\S}$ fields because it is not a field strength multiplet. Here duality helps us determine the observables because we know this model is equivalent to a K\"ahler model with a $P$ field. In that picture $\S,\ \tilde{\S}$ are possible observables but not $Y$. 
Eliminating $Y$ from \eqref{eqn:ysigmacoulomb2}\ and~\eqref{eqn:ysigmacoulomb3}\ gives the relation:
\begin{equation}
    -\tilde{k} n \log \left( \frac{\Sigma}{\mu} \right) + k \tilde{n} \log \left( \frac{\tilde{\Sigma}}{\mu} \right) = -i \tilde{k} t + i k \tilde{t} \quad\Rightarrow\quad \Sigma^{-\tilde{k} n} \tilde{\Sigma}^{k \tilde{n}} = \mu^{- \tilde{k} n + k \tilde{n}} e^{- i \tilde{k} t + i k \tilde{t}}.
\end{equation}
We can then use \eqref{eqn:ysigmacoulomb1} to express this relation in terms of a single $\Sigma$,
\begin{equation} \label{doubletrumpetring}
    \left(- \frac{k}{\tilde{k}} \right)^{k \tilde{n}} \Sigma^{- \tilde{k} n + k \tilde{n}} = \mu^{- \tilde{k} n + k \tilde{n}} e^{-i \tilde{k} t + i k \tilde{t}}.
\end{equation}
This is the quantum cohomology ring for the double trumpet model.

We can ask whether this ring is fundamentally different from the ring one would find in a toric case with no $Y$ couplings. While the basis of gauge symmetries we used for this model is the most convenient to see compactness of the $Y$ interval, it is not the best basis to understand the ring. We saw in section \ref{sect:doubletrumpetkahler} that we can change the basis of gauge fields for this model in order to transform it into the dual of a squashed toric model with $n$ fields of charge $-\tilde{k}/\h{k}$ and $\tilde{n}$ fields of charge $k/\h{k}$, where $\h{k}=\gcd(k,-\tilde{k})$. Note all these charges are positive. 

In this basis, $Y$ is only coupled to one field strength multiplet, which is precisely the combination of the original gauge field multiplets appearing in \eqref{eqn:ysigmacoulomb1}. That multiplet will be set to zero by the $Y$ equation of motion. The other field strength multiplet will obey a condition that is a function only of its FI parameter and the charges of the fields integrated out. The ring we constructed above in~\C{doubletrumpetring} is therefore the quantum cohomology ring of a weighted projective space.

We can apply the same reasoning to any model which is either a squashed K\"ahler model or a dual description of a squashed K\"ahler model. This is the case because squashing is a modification which only affects the $D$-terms, but not the $F$-terms which determine the quantum cohomology ring. We can therefore turn squashing off without changing the resulting ring.

\subsection{Exponential models}

The model in section \ref{sect:expysigmaoneu1} does not have Coulomb branch vacua because the $e^Y\Sigma$ coupling with only one $U(1)$ factor always sets $\Sigma=0$.

The Coulomb branch vacuum structure for the model of section \ref{sect:expysigmaseveralu1}\ is determined from the critical points of the effective twisted chiral superpotential:
\begin{align}
    S_{\tilde{W}}=\frac{1}{8\pi} \int d^2x d\theta^+ d\bar{\theta}^- \Bigg[& \Sigma \left(n \log \left(\frac{\Sigma}{\mu} \right) - n - k Y - i t(\mu)\right) + \tilde{\Sigma} \left(\tilde{n}\log \left(\frac{\tilde{\Sigma}}{\mu} \right) - \tilde{n} - \tilde{k} Y - i \tilde{t}(\mu)\right) \nonumber \\
    &+ \Sigma' \left(n'\log \left(\frac{\Sigma'}{\mu} \right) - n' - \kappa e^Y - i t'(\mu)\right) \Bigg].
\end{align}
These critical points satisfy the equations,
\begin{subequations} \label{eqn:coulombexpy}
\begin{align}
k\Sigma + \tilde{k}\tilde{\Sigma}+ \kappa e^Y\Sigma'&=0, \label{eqn:coulombexpy0} \\
n\log \left(\frac{\Sigma}{\mu} \right) -kY&=it(\mu), \label{eqn:coulombexpy1} \\
\tilde{n}\log \left(\frac{\tilde{\Sigma}}{\mu} \right) -\tilde{k}Y&=i\tilde{t}(\mu), \label{eqn:coulombexpy2} \\
n' \log\left(\frac{\Sigma'}{\mu} \right) - \kappa e^Y &=it'(\mu).\label{eqn:coulombexpy3}
\end{align}
\end{subequations}
Solving the three latter equations leads to $\Sigma$ solutions for any value of $Y$,
\begin{align} \label{eqn:sigmavacexpcouplings}
\left(\frac{\Sigma}{\mu}\right)^n &= e^{it(\mu)+kY}, & \left(\frac{\tilde{\Sigma}}{\mu}\right)^{\tilde{n}}&= e^{i\tilde{t}(\mu)+\tilde{k}Y}, & \left(\frac{\Sigma'}{\mu}\right)^{n'} &=  e^{it'(\mu)+\kappa e^Y}.
\end{align}
Each of these equations has a finite number of solutions for a fixed $Y$, giving a total of $n\tilde{n}n'$ vacua. 

Just like the case considered in~\eqref{eqn:cpncoulombvacuum}, it is valid to integrate out the charged fields as long as the right-hand sides of the equations appearing in~\eqref{eqn:sigmavacexpcouplings} are large. In equation~\eqref{eqn:coulombexpy1}, the FI parameter $t$ will run according to~\eqref{eqn:firunning},
\begin{equation}
    i t(\mu) = it (\Lambda) - n \log\frac{\mu}{\Lambda},
\end{equation}
and therefore $\Sigma / \mu$ will scale with $1/\mu$ as $\mu$ becomes smaller. Equivalently, solutions for $\Sigma$ will be independent of $\mu$. The $n$-dependence dropped out of this argument; it also drops out of analogous arguments which apply to equations~\eqref{eqn:coulombexpy2}\ and \eqref{eqn:coulombexpy3}. This implies that once we plug in $\Sigma$ into \eqref{eqn:coulombexpy0}, it can also be solved for $Y$ independently of $\mu$. The masses of the fields we integrated out will therefore become large when compared to $\mu$ for all vacuum solutions by the same argument found in section~\ref{sect:coulombvacuaderiv}.

We still have to solve the first equation,~\eqref{eqn:coulombexpy0}. To count the number of solutions, it is easier to work with the single-valued field $X=e^Y$. In terms of $X$, the remaining vacuum equation takes the form:
\begin{equation} \label{Xequation}
k e^{it/n} X^{k/n}+ \tilde{k} e^{i\tilde{t}/\tilde{n}} X^{\tilde{k}/\tilde{n}}+\kappa X e^{it'/n'+\kappa X/n'}=0.
\end{equation}
This equation has no dependence on the scale $\mu$. This can be seen by noting that $\mu$ drops out of equations~\eqref{eqn:coulombexpy}\ and~\eqref{eqn:sigmavacexpcouplings}. So one can use $t, \tilde{t}$ and $t'$ defined at the scale $\Lambda$ in equation~\C{Xequation}.  As a complex function of $X$, the left-hand side of this equation has an infinite number of zeroes, so we have infinite distinct vacua in the Coulomb branch. Since the masses of the integrated out fields are large in the IR, there is no obvious problem with this analysis. 


The structure we have found here is quite surprising and quite different from usual computations of quantum cohomology. It might well be indicative of a more generic vacuum structure found when examining generalized K\"ahler spaces beyond the lamp post. There are a couple of points to summarize: first, the Higgs branch geometry is compact for this model. In fact, the condition for the space to remain compact found in~\C{eqn:expysigmanointersection}\ can easily be preserved under RG flow. However, the space is non-K\"ahler and the structure of instanton corrections, which usually generate quantum cohomology, has yet to be understood in any detail. Similar comments apply to the observables of the theory. Because there is a non-zero $H$, the instanton configurations are likely to be complex field configurations. 

What we see is an infinite number of Coulomb branch vacua for this model. It is possible that further quantum corrections will lift these vacua, but since these vacua are seen from a holomorphic superpotential, it is not clear from where these quantum corrections might originate. One possibility are strong interactions between $Y$ and $\S$ generating an anomalous dimension for operators like $e^Y$. If we want to interpret the Coulomb branch as an operator ring, capturing quantum corrections to a classical ring of observables associated to the Higgs branch, then we note that $X=e^Y$ must be retained as an operator. The four operators $(\S, \tilde{\Sigma}, \Sigma', X)$ then satisfy the ring relations~\eqref{eqn:sigmavacexpcouplings} and~\C{Xequation}. 

One other possibility is that the Coulomb branch of this model should not be viewed as purely encoding data interpretable in terms of Higgs branch physics. In conventional GLSM examples, $\S$ fields can be related to Higgs branch fields via equations of motion. In this case, this is still true for the $\S$ fields but the neutral $Y$ field does appear on both branches, which perhaps suggests that the Coulomb branch might be viewed as distinct from the Higgs branch.

\subsection{An explicit example} \label{sect:explicitexample}

\begin{table}
\begin{center}
\begin{tabular}{l c c c}
\toprule
& $U(1)$ & $U(1)_s$ & $U(1)'$ \\ \midrule
$\Phi^i \ (\times n)$  & 1 & 0 & 0 \\
$\Phi_s \ (\times 1)$ & 1 & 1 & 0 \\
$\Phi'^j \ (\times n')$ & 0 & 0 & 1 \\
\bottomrule
\end{tabular}
\caption{\textit{The charge matrix for the example of section~\ref{sect:explicitexample}.}}
\label{tab:exponentialexample}
\end{center}
\end{table}

Let us remove some of the notational clutter to better understand and interpret what might be going on. Take 
the case $k=-\tilde{k}=1$. Additionally, take $\tilde{n}=1$. We  also want a better feel for how the exponential coupling with coefficient $\kappa$ is so dramatically changing the vacuum structure. So we will change basis for $(\S, \tilde{\Sigma})$, as outlined in section~\ref{sect:doubletrumpetring}, to 
$(\Sigma,\Sigma_s) = (\Sigma,\tilde{\Sigma}-\Sigma)$. This is a basis in which the $U(1)$ factors transparently describe the dual of a squashed $\mathbb{C}P^n$ model. We have $n$ fields $\Phi^i$ with charge 1 under the first $U(1)$ factor $\S$, which has no $Y$ coupling; there is one field $\Phi_s$ with charge $1$ under $\S$ and also charge 1 under $\S_s$. The $\S_s$ gauge symmetry has a $Y$ coupling. The phase of the $\Phi_s$ field is squashed in the $P$ picture. Finally there are fields $\Phi'^j$ with charge $1$ under $\S'$. This collection of fields appears in table~\ref{tab:exponentialexample}.

Supersymmetric vacua are determined by equations~\eqref{eqn:coulombexpy} which become
\begin{subequations} \label{eqn:coulombexpyexample}
\begin{align}
- \Sigma_s+ \kappa e^Y\Sigma'&=0, \label{eqn:coulombexpyexample0} \\
n \log \left( \frac{\Sigma}{\mu} \right) + \log\left( \frac{\Sigma + \Sigma_s}{\mu} \right) &= i t + i \tilde{t}, \label{eqn:coulombexpyexample1} \\
\log \left(\frac{\Sigma + \Sigma_s}{\mu}\right) + Y &= i \tilde{t}, \label{eqn:coulombexpyexample2} \\
n' \log \left( \frac{\Sigma'}{\mu} \right) - \kappa e^Y &= i t'.\label{eqn:coulombexpyexample3}
\end{align}
\end{subequations}
The first thing we would like to recover is the ring of the dual projective space, which should emerge in the limit $\kappa\rightarrow 0$. Setting $\kappa=0$ forces $\S_s=0$ from \eqref{eqn:coulombexpyexample0}. The remaining equations then decouple and in terms of $X=e^Y$ we find:
\be\label{basicring}
\S^{n+1} = \mu^{n+1}e^{i(t+\tilde{t})}, \qquad \S'^{\, n'} = \mu^{n'} e^{it'}, \qquad \S X = \mu e^{i\tilde{t}}. 
\ee
The first two relations are the familiar ones we expect for the $\mathbb{C}P^n \times \mathbb{C}P^{n'-1}$ model. The last relation constrains the operator $X$ in terms of $\S$. 

Now we turn on $\kappa$. It still seems natural to use \eqref{eqn:coulombexpyexample0}\ to solve for $\Sigma_s$ with $\Sigma_s = \kappa X \Sigma'$. The rest of the equations give the relations,
\begin{equation}
    \Sigma^n(\Sigma + \kappa X \Sigma') = \mu^{n+1} e^{i(t + \tilde{t})}, \qquad \Sigma'^{n'} e^{-\kappa X} = \mu^{n'} e^{it'}, \qquad X\left( \Sigma  + \kappa X\Sigma' \right) = \mu e^{i\tilde{t}}, 
\end{equation}
which are an intriguing deformation of the ring relations~\C{basicring}. Specifically, $\S$ and $\S'$ are now coupled as we might expect from figure~\ref{figure5}. This appears to be the case even in the classical limit where $r, \tilde{r}, r' \rightarrow \infty$, where the ring should correspond to a geometric ring of the generalized K\"ahler space.

Once we take $\kappa \neq 0$, the number of solutions to the equation determining $X$,
\begin{equation} \label{Xexplicitequation}
e^{it/n} X^{1/n} - e^{i\tilde{t}} X^{-1}+\kappa X e^{it'/n'+\kappa X/n'}=0,
\end{equation}
moves from finite to infinite. It is worth noting that if we try to perturbatively expand the zero solutions around $\kappa = 0$ to any finite order in $\kappa$, the number of solutions will still be finite. There are an infinite number of solutions that are not analytic around $\kappa = 0$. To make this more natural, we can examine a toy model with only one chiral superfield $X$ and a superpotential of the form
\begin{equation}
    W = X - e^{\kappa X}.
\end{equation}
Critical points of this superpotential obey the condition
\begin{equation}
    1 - \kappa e^{\kappa X} = 0.
\end{equation}
When $\kappa = 0$, this equation has no solutions. However, if $\kappa > 0$, the solutions are given by
\begin{equation}
    X = \frac{1}{\kappa}\log\left(\frac{1}{\kappa}\right).
\end{equation}
There is an infinite set of solutions, one for each branch of the logarithm. They are all non-analytic in $\kappa$, moving to infinite $|X|$ as $\kappa$ is taken to 0.

It would be very interesting to calculate elliptic genera for this class of models, in order to better understand the infinity of vacua we have found. Unfortunately, to compute the elliptic genus in a straightforward way, we need both $U(1)_L$ and $U(1)_R$ R-symmetries to be unbroken, which is not true for these compact models. As we have already discussed, it is challenging to find torsional examples which are both compact and conformal in this $(2,2)$ setting. Such models are possible with $(0,2)$ worldsheet supersymmetry. For non-compact models, the elliptic genus should generically have a non-holomorphic dependence on the torus modular parameter; see, for example~\cite{Gupta:2017bcp,Gupta:2018krl}. 

\subsection{Unified structure}

This Coulomb branch analysis can also be generalized to the unified case described in section \ref{sect:fysigma}. We allow for several fields $Y^\mu$, and any number of vector superfields $\Sigma_a$, coupled by twisted superpotential couplings of the more general form $f_a(Y)\Sigma_a$. We take the set of charged fields to consist of $n_a$ chiral superfields which are charged with charge $1$ only under one gauge symmetry corresponding to $\Sigma_a$. If we integrate out the charged superfields, we obtain an effective twisted superpotential,
\begin{equation}
    S_{\tilde{W}} = \frac{1}{8\pi}\int d^2x d\theta^+ d\bar{\theta}^- \sum_a \Sigma_a\left(n_a\left[\log\left(\frac{\Sigma_a}{\mu}\right)-1\right]-f_a(Y)-it_a(\mu)\right).
\end{equation}
Varying the superpotential with respect to $\Sigma_a$ gives equations of the form
\begin{equation} \label{eqn:unifiedstructurecoulombsigma}
n_a \log\left(\frac{\Sigma_a}{\mu}\right) - f_a = i t_a \quad \Rightarrow \quad  \left( \frac{\Sigma_a}{\mu} \right)^{n_a} = e^{it_a+f_a}.
\end{equation}
Varying with respect to $Y^\mu$ gives the conditions
\begin{align} 
\sum_a \del_\mu f_a \Sigma_a = 0.  
\end{align}
If we are interested in solving for vacua rather than studying rings, we can further substitute solutions to~\eqref{eqn:unifiedstructurecoulombsigma}\ giving:
\be \label{eqn:generalcoulombvacua}
\sum_a \del_\mu f_a e^{it_a/n_a+f_a/n_a}= \p_\mu \left( \sum_a n_a e^{it_a/n_a+f_a/n_a}\right)=0. 
\ee
As in the previous section, the masses of the fields that were integrated out become arbitrarily large as we flow to lower energies.

We now want to explore the generic number of solutions to~\eqref{eqn:generalcoulombvacua}. For simplicity, take one $Y$ field, and consider~\eqref{eqn:generalcoulombvacua} initially as a complex function of the cylinder variable $y$. For compactness $a>1$. We want to characterize the number of zeros of this function,
\begin{equation}\label{eqn:hX}
    h(y) = \sum_a \del_y f_a e^{it_a/n_a+f_a/n_a}. 
\end{equation}
The functions $f_a$ take the form 
$$
f_a = k_a y + {\tilde f}_a, \qquad  {\tilde f}_a= \sum_{m=-\infty}^\infty c_a^m e^{m y}, 
$$
for complex constants $c_a^m$. Some $k_a$ might be negative. We would like to move from the cylinder variable $y \sim y + 2\pi i$ to a single-valued variable. If it were not for the $n_a$ factors appearing in~\C{eqn:hX}, we would simply use $x=e^y$. Instead define $n_{\rm lcm} = {\rm lcm}\{n_a\}.$ We can then view $h(y)$ as a complex function of $z = e^{y/{n_{\rm lcm}}}$ rather than $y$. The penalty for this change of variable is that the equation $h(y)=0$ is replaced by a finite collection of equations in $z$ obtained by repeatedly shifting $y\rightarrow y + 2\pi i$. 

To proceed, we need to be able to say something about $f_a$. Usually, we do not want singular couplings in the classical Lagrangian so let us assume that $f_a$ is smooth with no singularities for finite values of $y$. Viewed as a function of $z$, this implies $f_a$ is holomorphic in $z$ away from $0$ and $\infty$. It is not particularly strange to also insist that ${\tilde f}_a$ is holomorphic in $z$. At least under this restriction, we can say something more about the number of zeros because $h(z)$ is analytic in the complex plane. Any analytic function with a finite number of zeros can be written in the form $P(z) e^{g(z)}$ where $g(z)$ is also an analytic function. Our $h(z)$ takes the form, 
\be
h(z) = \sum_a \left(k_a + \partial {\tilde f}_a(z) \right) z^\frac{k_a n_{\rm lcm}}{n_a} e^\frac{\tilde{f}_a(z)}{n_a} e^\frac{it_a}{n_a}, \qquad  {\tilde f}_a= \sum_{m=0}^\infty c_a^m z^{m \cdot n_{\rm lcm}}. 
\ee
If all ${\tilde f}_a(z)$ are identical then $h(z)$ can admit a finite number of zero solutions. Otherwise, we generically expect an infinite number of solutions as we saw in the example of section~\ref{sect:explicitexample}.

\section*{Acknowledgements}

It is our pleasure to thank L.~Anderson, J.~Gray, J.~Halverson, C.~Long, and  W.~Taylor for discussions about the degeneracy and fibration structure of currently known Calabi-Yau constructions. We would like to thank M.~Dine, D.~Kutasov, M.~R.~Plesser, and E.~Sharpe for helpful discussions. We are particularly grateful to I.~Melnikov for very helpful comments about these constructions.  J.~C. and S.~S. are supported in part by NSF Grant No. PHY-1720480.


\newpage

\appendix

\section{\texorpdfstring{$(2,2)$}{(2,2)} Superspace Conventions} \label{conventions}
\subsection{Basic conventions} \label{basicconventions}

In this Appendix, we state our conventions. 
We use superspace coordinates, 
$$(x^+,x^-,\theta^\pm,\bar{\theta}^\pm),$$ 
where the metric is Lorentzian, $x^\pm=\half (x^0\pm x^1)$ and $\del_\pm=\del_0\pm \del_1$ so that $\del_\pm x^\pm=1,\, \del_\pm x^\mp=0$. The superspace integration measure $d^4\theta=d\bar{\theta}^+ d\theta^+ d\bar{\theta}^- d\theta^-$ is defined so that $\int d^4\theta \, \theta^- \bar{\theta}^- \theta^+ \bar{\theta}^+=1$. The Levi-Civita tensor is defined by $\epsilon^{01}=1$ so $\epsilon^{-+}=\half$. The supersymmetry charges and super-derivatives can be written in terms of superspace coordinates as follows:
\begin{align}
Q_\pm&=\del_{\theta^\pm}+i\bar{\theta}^\pm\del_\pm, &\bar{Q}_\pm&=-\bar{\del}_{\theta^\pm}-i\theta^\pm\del_\pm, \\
D_\pm&=\del_{\theta^\pm}-i\bar{\theta}^\pm\del_\pm, &\bar{D}_\pm&=-\bar{\del}_{\theta^\pm}+i\theta^\pm\del_\pm.
\end{align}
These operators satisfy the algebras
\begin{align}
\{Q_\pm,\bar{Q}_\pm \}&=-2i\del_\pm, & \{D_\pm,\bar{D}_\pm \}&=2i\del_\pm.
\end{align}

A $(2,2)$ GLSM is constructed from a collection of constrained superfields. The first ingredient we use in this work are chiral superfields, $\Phi$, which  satisfy $\bar{D}_\pm \Phi=0$. They contain the following component fields:
\begin{equation}
\Phi=\phi+\sqrt{2}\theta^+\psi_+ +\sqrt{2}\theta^-\psi_-+2\theta^-\theta^+F+\dots,
\end{equation}
with all other terms involving derivatives of these fields. We also use twisted chiral superfields, $Y$, satisfying the conditions $\bar{D}_+ Y=D_-Y=0$. The $\theta$-expansion of these superfields takes the form:
\begin{equation}
Y=y+\sqrt{2}\theta^+\chi_++\sqrt{2}\bar{\theta}^-\chi_-+2\bar{\theta}^-\theta^+G+\dots.
\end{equation}

To gauge an abelian global symmetry acting on chiral superfields, we introduce a $U(1)$ vector superfield $V$, which is a real superfield. In Wess-Zumino gauge, the vector superfield has the expansion:
\begin{align}
V=& \ \theta^+\bar{\theta}^+ A_++\theta^-\bar{\theta}^-A_--\theta^-\bar{\theta}^+\sigma+\bar{\theta}^-\theta^+\bar{\sigma}+\sqrt{2}\theta^-\theta^+\bar{\theta}^+\lambda_+-\sqrt{2}\bar{\theta}^-\theta^+\bar{\theta}^+\bar{\lambda}_+\nonumber\\
&+\sqrt{2}\theta^-\bar{\theta}^-\bar{\theta}^+\lambda_--\sqrt{2}\theta^-\bar{\theta}^-\theta^+\bar{\lambda}_-+2\theta^-\bar{\theta}^-\theta^+\bar{\theta}^+D. 
\end{align}
A gauge transformation acts by sending,
\begin{align}
V &\rightarrow V+\frac{i}{2}(\bar{\Lambda}-\Lambda), & \Phi &\rightarrow e^{iQ\Lambda}\Phi, 
\end{align}
where the chiral superfield $\Phi$ has charge $Q$. From $V$ we can  build the field strength superfield
\begin{equation}\label{fieldstrength}
\Sigma=\bar{D}_+D_- V=\sigma+\sqrt{2}\theta^+\lambda_++\sqrt{2}\bar{\theta}^-\lambda_-+\bar{\theta}^-\theta^+(-2D+iF_{-+})+\dots,
\end{equation}
which is gauge-invariant and twisted chiral by construction. Similarly, if we want to gauge a global symmetry acting on twisted chiral superfields, we need to introduce a chiral vector superfield $\h{V}$.

Armed with these ingredients, we can describe the basic couplings of a $(2,2)$ GLSM. The canonical kinetic term of a chiral field $\Phi$ is given by
\begin{align} \label{chiralkinetic}
S&=\frac{1}{16\pi}\int d^2x d^4\theta \, \bar{\Phi} e^{2QV}  \Phi,	\nonumber \\
&=\frac{1}{4\pi}\int d^2x \big[-|\mathcal{D}_\mu \phi|^2+i\bar{\psi}_+\mathcal{D}_-\psi_++i\bar{\psi}_-\mathcal{D}_+\psi_-+|F|^2+QD|\phi|^2-Q^2|\sigma|^2|\phi|^2  \\
&\hspace{70pt}+Q\lambda_-\bar{\phi}\psi_+ +Q\bar{\psi}_+\phi\bar{\lambda}_-+Q\lambda_+\phi\bar{\psi}_-+Q\psi_-\bar{\phi}\bar{\lambda}_++Q\psi_+\sigma\bar{\psi}_-+Q\psi_-\bar{\sigma}\bar{\psi}_+\big]. \nonumber
\end{align}
Similarly canonical kinetic terms for a neutral twisted chiral superfield are given by
\begin{align}
S&=-\frac{1}{16\pi b}\int d^2x d^4\theta \, \bar{Y}Y, \nonumber \\
&=\frac{1}{4\pi b}\int d^2x \left[-|\mathcal{\del}_\mu y|^2+i\bar{\chi}_+\mathcal{\del}_-\chi_++i\bar{\chi}_-\mathcal{\del}_+\chi_-+|G|^2\right],
\end{align}
and finally the kinetic terms for a shift-charged chiral superfield are
\begin{align}
S&=\frac{b}{32\pi}\int d^2x d^4\theta (P+\bar{P}+2QV)^2, \nonumber \\
&=\frac{b}{4\pi} \int d^2x \big[-|\mathcal{D}_\mu p|^2+i\bar{\psi}_+\del_-\psi_++i\bar{\psi}_-\del_+\psi_-+|F|^2+QD(p+\bar{p})-Q^2|\sigma|^2\nonumber \\
&\hspace{70pt}+Q\lambda_- \psi_++Q\bar{\psi}_+\bar{\lambda}_-+Q\lambda_+\bar{\psi}_-+Q\psi_-\bar{\lambda}_+\big].
\end{align}
Kinetic terms for the gauge-field are built from $\S$, 
\begin{align}
S&=-\frac{1}{8e^2} \int d^2 x d^4\theta \, \bar{\Sigma}\Sigma, \nonumber \\
&=\frac{1}{2e^2}\int d^2x \left[-|\del_\mu\sigma|^2+i\bar{\lambda}_-\del_+\lambda_-+i\bar{\lambda}_+\del_-\lambda_++D^2-\half F_{\mu\nu}F^{\mu\nu}\right].
\end{align}
Superpotential couplings involve holomorphic combinations of chiral fields,
\begin{align}
S_W&=\frac{1}{8\pi}\int d^2x d\theta^+ d\theta^- \, W(\Phi)+\text{c.c.}, \nonumber \\
&=\frac{1}{4\pi}\int d^2x \left[\del_i W(\phi) F^i+\del_{ij}W\psi^i_+\psi^j_-\right]+ \text{c.c.}.
\end{align}

For holomorphic combinations of twisted chiral superfields, we can build an analogous twisted chiral superpotential ${\widetilde W}$. Specific examples central to our discussion involve the gauge-field strength $\S$ and a $Y$ field, taking the form
\begin{align}
S_{\widetilde{W}} &=-\frac{k}{8\pi}\int d^2x d\theta^+ d\bar{\theta}^- \, Y\Sigma +\text{c.c.}, \nonumber \\
&=\frac{k}{4\pi}\int d^2x\left[2\rept (y) D+\impt (y)F_{-+}-(\sigma G+\chi_+\lambda_-+\lambda_+\chi_-+\text{c.c.})\right], \label{eqn:ysigmacomponents}
\end{align}
or
\begin{align} \label{eqn:expysigmacomponents}
S_{\widetilde{W}} &=-\frac{\kappa}{8\pi}\int d^2x d\theta^+ d\bar{\theta}^- \, e^Y\Sigma +\text{c.c.},  \\
&=\frac{\kappa}{4\pi}\int d^2x\left[2\rept (e^y) D+\impt (e^y)F_{-+}-(e^y(\sigma G+\chi_+\lambda_-+\lambda_+\chi_-+\sigma\chi_+\chi_-)+\text{c.c.})\right]. \nonumber 
\end{align}
Note that compatibility of these couplings with the periodicity $Y \sim Y + 2\pi i$ imposes the constraint $k \in \mathbb{Z}$, but there is no such restriction on $\kappa$. A more basic example is the Fayet-Iliopoulos (FI) coupling for an abelian gauge-field, given by
\begin{equation}\label{FIrtheta}
S_{FI}=-\frac{it}{8\pi}\int d^2x d\theta^+d\bar{\theta}^-\,\Sigma+\text{c.c.}=\frac{1}{4\pi}\int d^2x\left[-rD+\theta\epsilon^{\mu\nu}F_{\mu\nu}\right], 
\end{equation}
where
\begin{equation}
t=\frac{ir}{2}+\theta.
\end{equation}

\section{Dualization} \label{sect:Tdualityexpanded}

Here we summarize the various useful $(2,2)$ dual descriptions found, for example, in~\cite{Hori:2000kt}. We will take special care to ensure the correct normalization for the Lagrange multiplier terms so that circles in both the original and dual descriptions have $2\pi$ periodicity.  

\subsection{Component picture and periodicity} \label{sect:dualityperiodicity}

We will derive T-duality by proving equality of the path integrals over the two dual theories, through the construction of a theory with auxiliary fields that can be transformed into either of the dual theories. Schematically, we want to prove equivalence between a theory with action
\begin{equation} \label{eq:tdualityscalar}
    S = - \frac{b}{4\pi} \int d^2x \del_\mu \varphi \del^\mu \varphi
\end{equation}
and one given by
\begin{equation} \label{eq:tdualscalar}
    S = - \frac{1}{4\pi b} \int d^2x \del_\mu \theta \del^\mu \theta,
\end{equation}
where both $\varphi$ and $\theta$ are $2\pi$-periodic real fields. The strategy will be to rewrite the first theory replacing $d\varphi$ with an unconstrained one-form, adding the right Lagrange multiplier so that the theories are equivalent.

On a topologically non-trivial worldsheet whose first cohomology group has rank $n$, take $\{ \omega_i \}, \, i=1,\dots,n$ to be a basis for the closed non-trivial one-forms dual to a basis of non-trivial cycles such that the matrix $\int \omega^i \wedge \omega^j=J^{ij}$ is an element of $SL(n,\mathbb{Z})$. A generic closed one-form can then be written
\begin{equation} \label{eq:closedoneform}
c=c_\mu dx^\mu = d\varphi_0+ \sum_i a_i \omega^i,
\end{equation}
with $\varphi_0$ single-valued and $a_i$ real numbers.

To dualize $\varphi$ in \eqref{eq:tdualityscalar}, we will substitute $\del_\mu \varphi \rightarrow c_\mu$ and add a Lagrange multiplier $\theta$ to enforce $dc=0$. The action then becomes
\begin{equation} \label{eq:tdualitys0}
S_0 = - \frac{b}{4\pi} \int d^2x c_\mu c^\mu + \kappa \int \theta dc
\end{equation}
where $\kappa$ is a multiplicative constant. When we integrate out $\theta$, the equation $dc=0$ allows us to trivialize $c=d\varphi$ and return to the original theory. However, as in \eqref{eq:closedoneform}, $\varphi$ in this expression will not be a single-valued field if the worldsheet is topologically non-trivial, so we need to find the relation between the periodicities of $\theta$ and $\varphi$ induced by this coupling.

Defining the Lagrange multiplier $\theta$ to have period $T_\theta$, we can expand $d\theta$ as
\begin{equation}
d\theta=\del_\mu\theta d x^\mu =d\theta_0+T_\theta \sum_i n_i \omega^i,
\end{equation}
with single-valued $\theta_0$ and integers $n_i$. We then have
\begin{equation}
 \int c \wedge d\theta=\int c_\mu \del_\nu\theta dx^\mu \wedge dx^\nu =\int d^2 x \epsilon^{\mu\nu} c_\mu \del_\nu \theta =  T_\theta\sum_{i,j} a_i J^{ij} n_j,
\end{equation}
where we added the intermediate forms for later use. Now when we perform the path integral over $\theta$, the $\theta_0$ part gives $dc=0$, but the integral also includes a sum over $n_i$, which gives
\begin{equation}
\sum_{n_j} \exp \left(-i \kappa T_\theta a_i J^{ij} n_j\right) \propto \prod_i \sum_{m_i} \delta(\kappa T_\theta a_i-2\pi m_i),
\end{equation}
constraining $a_i$ to be integer multiples of $\frac{2\pi}{\kappa T_\theta}$, so comparing to \eqref{eq:closedoneform}, we see that $c=d\varphi$ where $\varphi$ is periodic with period  $T_\varphi=\frac{2\pi}{\kappa T_\theta}$. Therefore, if we want $T_\theta=T_\varphi=2\pi$, we should take $\kappa=1/(2\pi)$.

Now that we have fixed the constant in \eqref{eq:tdualitys0}, the dual action can be obtained by instead integrating out $c_\mu$. Since the action is quadratic, the path integral sets $c_\mu$ to the solution of its classical equation of motion,
\begin{equation}
    b c^\mu = - \epsilon^{\mu\nu} \del_\nu\theta,
\end{equation}
or in components
\begin{equation}
    b c_\pm = \pm \del_\pm \theta.
\end{equation}
The dual action we obtain is \eqref{eq:tdualscalar}.

\subsection{\texorpdfstring{$(2,2)$}{(2,2)} duality}

Consider now a free $(2,2)$ chiral field $P$ with periodicity $P \sim P + 2\pi i$:
\begin{equation}
S=\frac{b}{16\pi}\int d^2x d^4\theta \, |P|^2.
\end{equation}
To dualize the imaginary component of $P$, we will substitute $P +\bar{P}\rightarrow 2B$ where $B$ is a generic real superfield. Denoting the imaginary part of $P$ by $\varphi$ and the one-form in $B$ by $c_\mu$, we see this corresponds to the component substitution in the previous section since
\begin{align}
P+\bar{P} &= \dots - i\theta^+\bar{\theta}^+\del_+(p-\bar{p})-i\theta^- \bar{\theta}^- \del_-(p-\bar{p}),  \nonumber \\
&=\dots + 2 \theta^+ \bar{\theta}^+ \del_+ \varphi +2\theta^- \bar{\theta}^-\del_-\varphi,
\end{align}
and we expand
\begin{equation}
B = \dots + \theta^+ \bar{\theta}^+ c_+ + \theta^- \bar{\theta}^-c_-
\end{equation}
so $P +\bar{P}\rightarrow 2B$ is the $(2,2)$ extension of $d\varphi \rightarrow c$.

Carrying out the substitution and including a Lagrange multiplier $F$, the action becomes
\begin{align}
S_0&=\frac{1}{8\pi}\int d^2x d^4\theta \left[b B^2+F\bar{D}_+D_-B-\bar{F}D_+\bar{D}_-B\right],\nonumber \\
& = \frac{1}{8\pi}\int d^2x d^4\theta \left[b B^2-D_- \bar{D}_+ F B+ \bar{D}_- D_+ \bar{F} B\right],\nonumber \\
&=\frac{1}{8\pi}\int d^2x d^4\theta \left[b B^2-B(Y+\bar{Y})\right], \label{eqn:S0PtoY}
\end{align}
where we integrated by parts and defined $Y=D_- \bar{D}_+ F$ to obtain the second form of the action. Note that $Y$ is twisted chiral by definition. 

If we integrate out the Lagrange multiplier $F$, its equation of motion is solved by setting $B$ to the sum of a chiral and its conjugate and we recover the original theory. We should check that the factor multiplying the Lagrange multiplier term is the one determined in section \ref{sect:dualityperiodicity}: expand
\begin{equation}
Y + \bar{Y} = \dots +2\theta^+\bar{\theta}^+ \del_+ \theta - 2 \theta^-\bar{\theta}^-\del_-\theta,
\end{equation}
from which we see
\begin{align}
S_L&=-\frac{1}{8\pi}\int d^2x d^4\theta B (Y+\bar{Y}) \nonumber \\
&= - \frac{1}{4\pi}\int d^2x \left[c_-\del_+\theta - c_+\del_-\theta \right] + \dots \nonumber \\
&=- \frac{1}{2\pi} \int d^2x \epsilon^{\mu\nu}c_\mu\del_\nu\theta +\dots
\end{align}
Therefore the periods of $\theta$ and $\varphi$ will be related by $T_\theta T_\varphi = 4\pi^2$; if one period is $2\pi$ periodic, so will the other.

Solving~\eqref{eqn:S0PtoY} for $B$ instead, we find the dual action
\begin{equation}
S_d=-\frac{1}{32\pi b}\int d^2x d^4\theta (Y+\bar{Y})^2=-\frac{1}{16\pi b}\int d^2x d^4\theta |Y|^2.
\end{equation}
This is a canonical kinetic term for a twisted chiral. Note the duality has mapped
\begin{equation}
b(P+\bar{P}) = Y+\bar{Y},
\end{equation}
which for their scalar component fields becomes
\begin{align}
b(p+\bar{p}) & = y+\bar{y}, \\
b\del_\pm (p-\bar{p}) & = \pm \del_\pm (y-\bar{y}). \label{eqn:ypdualitymap}
\end{align}
This confirms our assertion that we are performing a T-duality transformation on the imaginary part of $p$. The imaginary part of $p$ parametrizes a circle with radius $\sqrt{b}$, while the circle parametrized by the imaginary part of $y$ has radius $1/\sqrt{b}$.

Such a field $P$ can be axially charged, making its imaginary part a two-dimensional Stueckelberg field. The action is simply
\begin{equation}
S=\frac{b}{32\pi}\int d^2xd^4\theta \left(P+\bar{P}+2V\right)^2.
\end{equation}
Using the same substitution as above gives,
\begin{equation}
S_0=\frac{1}{8\pi}\int d^2x d^4\theta\left[b (B+V)^2-B(Y+\bar{Y})\right],
\end{equation}
which we can solve for $B$ to find
\begin{align}
S_d&= \frac{1}{8\pi}\int d^2x d^4\theta \left[-\frac{1}{4 b} (Y+\bar{Y})^2+(Y+\bar{Y})V\right], \nonumber \\
&=-\frac{1}{16\pi b}\int d^2x d^4\theta  \bar{Y}Y-\frac{1}{8\pi}\int d^2x d\theta^+ d\bar{\theta}^- Y\Sigma + \text{c.c.}.
\end{align}
The last term in the action is the coupling between $Y$ and $\Sigma$ in~\eqref{eqn:twistedsuperpotential}. Note that a field $P$ with charge $Q_p$ is dual to $Y$ with coupling $k_y = Q_p$.
The duality maps are similar to~\eqref{eqn:ypdualitymap} except the $p$ side will now feature covariant derivatives $\mathcal{D}_\pm p = \del_\pm p + A_\pm$, so both sides of the maps are gauge-invariant.

Alternatively, if we start from a $(2,2)$ chiral $\Phi$ parametrizing a plane, we can redefine $\Phi=e^\Pi$ to dualize the phase of $\Phi$ by similar techniques. Since $(2,2)$ theories are not chiral, no non-trivial Jacobian results from this redefinition. Replacing $\Pi+\bar{\Pi}\rightarrow 2B$ gives,
\begin{equation}
S_0 =\frac{1}{16\pi}\int d^2xd^4\theta\, \left[e^{2B}-2B(Y+\bar{Y})\right],
\end{equation}
from which we can once again solve for $B$ to find
\begin{equation}
S_d=-\frac{1}{16\pi}\int d^2xd^4\theta \,(Y+\bar{Y})\log(Y+\bar{Y}).
\end{equation}
The same procedure can be performed if $\Phi$ is charged,
\begin{align}
S=\frac{1}{16\pi}\int d^2x d^4\theta \,|\Phi|^2 e^{2V} 
\quad \rightarrow\quad S_0 =\frac{1}{16\pi}\int d^2x d^4\theta \,\left[e^{2B+2V}-2B(Y+\bar{Y})\right], 
\end{align}
which gives the dual action:
\begin{align}
S_d &=-\frac{1}{16\pi}\int d^2x d^4\theta \,\left[(Y+\bar{Y})\log(Y+\bar{Y})-2 (Y+\bar{Y})V\right], \nonumber \\
&=-\frac{1}{16\pi}\int d^2xd^4\theta\, (Y+\bar{Y})\log(Y+\bar{Y}) -\frac{1}{8\pi}\int d^2x d\theta^+d\bar{\theta}^-\, Y\Sigma +\text{c.c.}.
\end{align}
In this case, for the theories to be quantum-mechanically equivalent, we must add a term to the twisted superpotential of the dual theory reflecting an instanton correction in the original theory~\cite{Hori:2000kt}. The full dual action is
\begin{equation}
S_d = -\frac{1}{16\pi}\int d^2xd^4\theta\, (Y+\bar{Y})\log(Y+\bar{Y}) -\frac{1}{8\pi}\int d^2x d\theta^+d\bar{\theta}^- \,\left[Y\Sigma + \mu e^{-Y}\right] +\text{c.c.}.
\end{equation}

\section{Central Charge and Anomaly Details}\label{app:centralchargedetails}
The leading singularities in the OPE of the basic fields of the model~\eqref{eq:generalPYmodel}\ are:
\begin{align}
\phi_i(x) \phi_j(0) &\sim  -\delta_{ij} \log \left( x^2 \right), & \psi_{i,\pm}(x) \bar \psi_{j,\pm}(0) &\sim -\frac{i\delta_{ij}}{x^{\pm\pm}}, \nonumber\\
p_\alpha(x) p_\beta(0) &\sim  - \frac{\delta_{\alpha \beta}}{b_\alpha} \log \left( x^2 \right), & \eta_{\alpha,\pm}(x) \bar \eta_{\beta,\pm}(0) &\sim -\frac{i\delta_{\alpha\beta}}{ b_\alpha x^{\pm\pm}}, \\
y_\mu(x) y_\nu(0) &\sim  -b_\mu \delta_{\mu\nu} \log \left( x^2 \right), & \chi_{\mu,\pm}(x) \bar \chi_{\nu,\pm}(0) &\sim -\frac{ib_\mu \delta_{\mu\nu}}{x^{\pm\pm}}, \nonumber\\
\sigma_a(x) \sigma_b(0) &\sim  -\frac{e^2_a}{2\pi}\delta_{ab} \log \left( x^2 \right), & \lambda_{a,\pm}(x) \bar \lambda_{b,\pm}(0) &\sim - \frac{e^2_a}{2\pi}\frac{i\delta_{ab}}{x^{\pm\pm}}. \nonumber
\end{align}
The bottom component of the classical supercurrent $\mathcal{J}_{--}^0$ includes the composite operator $\sum_i \psi_{i,-}\bar\psi_{i,-}(x)$, which we define via point-splitting
\begin{align}
\sum_i \psi_{i,-}\bar \psi_{i,-}(x) &:= \lim_{y\to x}\left( \sum_i \psi_{i,-}(y) e^{i\int_y^x Q_{ia}A_a }\bar \psi_{i,-}(x) - \frac{i}{\left(x-y\right)^{--}}\right) \\
& = :\sum_i \psi_{i,-}\bar \psi_{i,-}(x): -\sum_i Q_{ia}A_{a--}(x) -\sum_i Q_{ia}A_{a++}(x)\lim_{y\to x}\frac{\left(x-y\right)^{++}}{\left(x-y\right)^{--}} \nonumber.
\end{align}
The anomaly in the supercurrent is determined by this operator:
\begin{align}
\left. \bar D_- \mathcal{J}_{--}^0 \right| = -\frac{1}{4\pi} \com{\bar Q_+}{\sum_i \psi_{i,-}\bar\psi_{i,-}(x)} = -\frac{\sqrt{2}}{4\pi} \sum_i Q_{ia}\lambda_{a,-}(x) = \frac{\sum_i Q_{ia}}{4\pi} \left.  \bar D_- \Sigma_a \right|.
\end{align}
Thus, the anomaly
\begin{equation}
\gamma_a = \sum_i Q_{ia}.
\end{equation}

We determine the central charge of the $\mathcal{N} = 2$ Virasoro algebra generated by $\mathcal{J}_{--} + \mathcal{F}_{--}$ by considering the leading singularity of the current-current OPE. The $R$-current, the bottom component of the superfield is
\begin{align}
j_{--} & = \frac{i \alpha_i}{4\pi} \phi_i \mathcal{D}_{--} \bar \phi_i + \frac{i \gamma_\alpha}{4\pi} \left( \mathcal{D}_{--} p_\alpha - \mathcal{D}_{--} \bar p_{\alpha} \right) - \frac{i \gamma_\mu}{4\pi \sqrt{b_\mu}} \partial_{--}\left( y_\mu - \bar y_\mu \right) -\frac{i}{2 e_a^2} \sigma_a \partial_{--} \bar \sigma_a \\
& - \frac{ 1 - \alpha_i}{4\pi} \psi_{i,-} \bar\psi_{i,-} - \frac{b_\alpha}{4\pi} \eta_{\alpha,-} \bar\eta_{\alpha,-} + \frac{1}{4\pi b_\mu} \chi_{\mu, -}\bar \chi_{\mu, -}. \nonumber
\end{align}
Using the OPEs from above, we see that
\begin{align}
j_{--}(x) j_{--}(0) &\sim -\frac{\left( \sum_i \left( 1 - 2\alpha_i \right) - N_{U(1)} + N_P + N_Y + 2 \sum_\alpha \frac{\gamma_\alpha^2}{b_\alpha} + 2 \sum_\mu \frac{\gamma_\mu^2}{b_\mu} \right)}{(x^{--})^2} + \ldots, \\
&= -\frac{c/3}{(x^{--})^2} + \ldots. \nonumber
\end{align}
Hence the quoted formula in the main body of the text~\C{centralcharge}. 

\newpage
\bibliographystyle{utphys}
\bibliography{master}

\end{document}